\documentclass[12pt]{article}
\pdfoutput=1 
\usepackage{jheppub} 
\usepackage{color}
\usepackage{amsmath,amsfonts,amssymb}
\usepackage{graphicx,epsfig}
\usepackage{psfrag}
 \usepackage{makecell}
\usepackage{hyperref}

\usepackage[normalem]{ulem}
\usepackage{enumerate}
\usepackage{yfonts}

\usepackage{psfrag}

\usepackage{graphicx}
\usepackage{cancel}
\usepackage{slashed}

\usepackage[font=small,labelfont=bf]{caption}
\usepackage{subcaption}

\newcommand{\be}{\begin{equation}}
\newcommand{\bea}{\begin{eqnarray}}
\newcommand{\ee}{\end{equation}}
\newcommand{\eea}{\end{eqnarray}}

\title{ A review on analytical studies in Gravitational Lensing}

\author[a]{Abhishek Chowdhuri,}
\author[a]{Saptaswa Ghosh,}
\author[a]{Arpan Bhattacharyya }
\affiliation[a]{\it Indian Institute of Technology, Gandhinagar, Gujarat-382355, India}
\emailAdd{chowdhuri$\_$abhishek@iitgn.ac.in}
\emailAdd{saptaswaghosh@iitgn.ac.in}
\emailAdd{abhattacharyya@iitgn.ac.in}
\abstract{In this study, we review some current studies on Gravitational Lensing for black holes, mainly in the context of general relativity. We mainly focus on the analytical studies related to lensing with references to observational results. We start with reviewing lensing in spherically symmetric Schwarzschild spacetime, showing how to calculate deflection angles before moving to the rotating counterpart, the Kerr metric. Furthermore, we extend our studies for a particular class of newly proposed solutions called black-bounce spacetimes and discuss throughout the review how to explore lensing in these spacetimes and how the various parameters can be constrained using available astrophysical and cosmological data. }

\begin{document}
\maketitle
\flushbottom


\section{Introduction}
The importance of gravitational lensing began with Eddington's observation of the light deflection of the Sun\cite{Dyson:1920cwa}. The experiment proved to be one of the major milestones in favour of Einstein's theory of General Relativity (GR)\cite{Carroll:1997ar}. The theory has proved itself to be the most successful theory of classical gravity. However, GR has some hiccups in the form of singularities, not being able to understand the quantum theory of it, along with the ubiquitous nature of dark matter and dark energy, forcing us to conclude that maybe there's more to the story. Following this line of thought, efforts have been made to introduce modifications to the Einstein-Hilbert Lagrangian \cite{PhysRevD.68.104012,Kanti:1995vq,Horndeski:1974wa,Eling:2004dk,Moffat:2014aja,Gao:2014soa,Gao:2018izs,Gross:1986iv,Gross:1986mw,deRoo:1992zp,Lovelock:1971yv,Lovelock:1972vz}. However, taking in a cue from Eddington, gravitational lensing still remains of the most indispensable tools to check and put constraints on our theory parameters.

The theory behind gravitational lensing started when O. Lodge asked a simple question in \cite{Lodge} on the behaviour of light in a gravitational field. Interestingly, he concluded that, unlike a convex lens, the gravitational field does not focus light rays into a focal point. However, a logarithmically shaped concave lens can indeed mimic the effect of a spherically symmetric gravitational field onto the light. Inspired by such insights
  O. Chwolson came up with the possibility of observing ring-like images in cases of axial symmetry. Nowadays, these ring-like images are called
Einstein rings \cite{Ch1,Ch2}. Remarkably such images were indeed observed by J. Hewitt et al. using the Very Large Array, a collection of
radio telescopes in the US. The radio source was MG1131+0456, and it was found when a background (radio) galaxy is distorted into an almost closed ring\cite{hewitt1,hewitt2,hewitt3}.
Later, such rings were also found in the infrared and even in the optical spectrum \cite{Hammer} 
and more recently by James Webb Telescope \cite{JST}. Typically, the diameter ranges between a few arcseconds or less. 
\par
The discovery of quasars and such rings lead to a rapid change in the field of gravitational lensing. We are now well aware of lensing techniques such as weak lensing, where the small deformations of many background galaxies are statistically evaluated for determining the (dark) matter in a foreground galaxy cluster, and microlensing, where the light curve of a star is registered that moves transversely to the line of sight behind a (dark and compact) mass. For a thorough exposition of gravitational lensing, including an overview of all pre-1992 observations, the reader may consult the monograph \cite{Falco}. More up-to-date information can be found in \cite{Perlick11}.
\par
In all the lensing observations mentioned above, the bending angles are so small that a weak-field approximation for the gravitational field is applicable. A much simpler quasi-Newtonian formalism has been used. This approximation centres around a so-called lens map or lens equation, which is very intuitive. This lens map of the weak-field formalism has proven extremely useful for evaluating the above-mentioned lensing phenomena. It is discussed in detail, e.g.~in \cite{Falco}. Additional material can be found, e.g., in the Living Review by Wambsganss\cite{Wamb}, which is completely based on this approximation formalism. However, there are astrophysical scenarios where the bending angles are not small. In such cases, the weak lensing theory is not valid anymore. The objective of this review is to give an overview of the various analytical techniques used to study strong lensing and to give some insights into the relation of lensing parameters to the gravity theory itself.\par

This expectation is nurtured mainly by the increasing evidence of a black hole at the centre of our galaxy. Black holes arise as solutions of GR and play an important role in testing various aspects of it. Recent observations coming from ``Event Horizon Telescope" (EHT) \cite{EventHorizonTelescope:2019dse,EventHorizonTelescope:2019uob,EventHorizonTelescope:2019jan, EventHorizonTelescope:2019ths,EventHorizonTelescope:2019pgp, EventHorizonTelescope:2019ggy, EventHorizonTelescope:2022xnr, EventHorizonTelescope:2022vjs, EventHorizonTelescope:2022wok, EventHorizonTelescope:2022exc, EventHorizonTelescope:2022urf, EventHorizonTelescope:2022xqj,
EventHorizonTelescope:2021dqv} collaboration and LIGO, Virgo and KAGRA Collaboration \cite{LIGOScientific:2016aoc,LIGOScientific:2016vlm, LIGOScientific:2016sjg, LIGOScientific:2017bnn,LIGOScientific:2021qlt,Kokeyama:2020dkg} have provided us with enough observational evidence that ensures the existence of supermassive black holes. Furthermore, these observations provide a unique opportunity to test various aspects of strong gravity \cite{LIGOScientific:2019fpa,LIGOScientific:2021sio,Psaltis:2018xkc,Himwich:2020msm, Gralla:2020pra,Vagnozzi:2022moj,Volkel:2020xlc,Bambi:2019tjh,Vagnozzi:2019apd,Allahyari:2019jqz,Khodadi:2020jij}. This necessitates us to go beyond weak-field approximation and consider lensing by the strong gravitational field, thereby motivating  us to consider the strong field limit of bending angle. It provides important information about the intrinsic parameters like spin, mass and charge of the black holes and the parameters of the underlying gravity theory. Hence its observational implications have been investigated in recent times  \cite{Cunha:2015yba,Wang:2017hjl,Younsi:2016azx,Bisnovatyi-Kogan:2018vxl, Banerjee:2019nnj,Tsupko:2019mfo,Tsupko:2019pzg, Mishra:2019trb, Vagnozzi:2020quf,Li:2020drn,Perlick:2017fio,Wei:2018xks,Chowdhuri:2020ipb,Papnoi:2014aaa,Lin:2022ksb,Chandrasekhar:579245,Adler:2022qtb,Virbhadra:2022ybp,Bambi:2019tjh,Vagnozzi:2019apd,Allahyari:2019jqz,Khodadi:2020jij,Roy:2021uye,Chen:2022nbb,Chen:2022kzv}\footnote{This list is by no means exhaustive. Interested readers are referred to this review \cite{Perlick:2010zh} and citations there for more details.}. An analytical description of the deflection angle for the spherically symmetric black hole in the strong gravitational field has been proposed in \cite{Bozza:2002zj}. This was further based on \cite{Virbhadra:1999nm, Frittelli:1999yf}. Later it was extended for rotating black holes in \cite{Bozza:2002af}. A lens equation has been derived in \cite{Virbhadra:1999nm}, which has been further generalized in \cite{Bozza:2008ev} where the underlying black hole serves as a lens. Numerous investigation of Einstein rings \cite{Einstein:1936llh}, and strong gravitational lensing based on the work has been made for various black hole spacetimes and compact objects \cite{Eiroa:2010wm,Bin-Nun:2010lws,Amarilla:2010zq,Stefanov:2010xz, Wei:2011nj,Chen:2011ef,Gyulchev:2012ty,Tsukamoto:2012xs,Sahu:2012er,Chen:2012kn,Wei:2013kza,Atamurotov:2013sca,Eiroa:2013nra,Wei:2014dka,Tsukamoto:2014dta,Liu:2015zou,Wei:2015qca,Sotani:2015ewa,Bisnovatyi-Kogan:2015dxa,Sharif:2015qfa,Younas:2015sva,Chen:2015cpa,Schee:2017hof,Man:2012ivp,Tsukamoto:2016jzh,Wang:2016paq,Tsukamoto:2016qro,Zhao:2016ltm,Aldi:2016ntn,Lu:2016gsf,Cavalcanti:2016mbe,Zhao:2016kft,Shaikh:2017zfl,Rahman:2018fgy,Virbhadra:2022iiy,Hsieh:2021scb,Hsieh:2021rru,Edery:2006hm,Ghosh:2022mka}\footnote{Again, this list is by no means exhaustive. Interested readers are referred to the references and citations of these papers.}. \par
Note that in this review, we will be focusing on the analytical computation of the deflection angle of light rays using null geodesic equations. But there are other methods of calculating deflection angle analytically, for instance, the \textit{material medium approach}. In this approach, one maps the problem effectively to a problem of light propagation through a medium with a particular refractive index determined by the strength gravitational field of the original spacetime for which one wants to calculate the deflection angle. For more details, interested readers are referred to some of these references \cite{Balazs:1958zz,PhysRev.118.1396,1969AJ.....74..320A,1971GReGr...2..347D,Mashhoon:1973zz,Mashhoon:1975ki,PhysRevD.22.2950,PhysRevD.67.064007,Ye:2007av,2010Ap.....53..560S,Roy:2014hca,Chakraborty:2014ava,Chakraborty:2015aua} and citations of these references.  
\par
Furthermore, in recent times, several static metrics known as \textit{Black Bounce metric} have been proposed in \cite{Simpson:2018tsi,Lobo:2020ffi, Mazza:2021rgq, Franzin:2021vnj}. What makes this spacetime interesting is that they have an extra parameter which regularises the central singularity unlike the usual black hole spacetime. The black-bounce spacetime interpolates between a black hole and (Traversable) wormhole metric depending on the choice of underlying regularisation parameters. When the solution does not admit any horizon, it corresponds to a  wormhole solution. Recently, the gravitational lensing in strong deflection limit for these black bounce spacetime has been studied \cite{Tsukamoto:2020bjm, Nascimento:2020ime, Guerrero:2021ues,Islam:2021ful, Ghosh:2022mka}. From the lensing perspective, this kind of spacetime provides us with extra tunable parameters; hence, it is interesting to investigate the effect of this extra parameter from the theoretical and observational points of view. In this review, besides discussing some analytical results about the computation of strong deflection angle in the background of Schwarzschild and Kerr-Newman black holes, we also discuss the effect of this black bounce regularization parameter on the computation of deflection angle and, finally, on the radius of the Einstein ring. \par

The article is organized as follows. In Sec.~(\ref{sec1}), we first review the deflection angle, the angular radius of Einstein rings and Shapiro time delay for a Schwarzschild spacetime. In Sec.~(\ref{sec4}), we review a general formalism of computation of the equatorial deflection angle for Kerr-Newman spacetime. Then, in Sec.~(\ref{sec5}), we discuss the computation of the deflection angle for the black bounce metric. We also review the computation of the Einstein ring radius of this case and comment on its dependence on charge and regularisation parameters and its observational implications. Finally, we give concluding remarks in Sec.~(\ref{sec9}). Some necessary details are given in Appendix~(\ref{appendixA}). Also, we have set the value of the speed of light $c$ and Newton's gravitational constant $G$ to unity.



 \section {Lensing for spherically symmetric Schwarzschild black hole}\label{sec1}
 In this section, we start by reviewing gravitational lensing in the simplest setup i.e for the Schwarzschild solution. The metric reads:
 \begin{equation}
     ds^{2}=-\Big(1-\frac{2M}{r}\Big)dt^{2}+\frac{dr^{2}}{1-\frac{2M}{r}}+r^{2}(d\theta^{2}+\sin^{2}\theta d\phi^{2})\,.
 \end{equation}
The Lagrangian for the particle $\Bar{L}=\frac{1}{2}g_{\mu\nu}(x)\dot{x}^{\mu}\dot{x}^{\nu}$ reads: 
\begin{equation}\label{1}
    \Bar{L}=\frac{1}{2}g_{\mu \nu}(x)\dot{x}^{\mu}\dot{x}^{\nu}=\frac{1}{2}\Big[-\Big(1-\frac{2M}{r}\Big)\dot{t}^{2}+\frac{\dot{r}^{2}}{1-\frac{2M}{r}}+r^{2}\dot{\phi}^{2}\Big]
\end{equation}
We are only considering geodesics on the equatorial plane where $\theta=\frac{\pi}{2}$. Hence, $\sin\theta=1$ along these geodesics and since $\theta$ is fixed every $\dot{\theta}$ component is $0$. Owing to the symmetry of the solution, we have two constants of motion which we can read off from the Euler-Lagrange equations.
The $t$-component of equation of motion gives,
\begin{equation} \label{rev1}
    \Big(1-\frac{2M}{r}\Big)\dot{t}=E=\textrm{constant}
\end{equation}
While the $\phi$ component gives,
\begin{equation} \label{rev2}
    r^{2}\dot{\phi}=L=\textrm{constant}
\end{equation}
In this article we will mainly focus on the lensing of light rays \footnote{For deflection of massive particle one needs to consider time like trajectories given by $g_{\mu\nu}\dot{x}^{\mu}\dot{x}^{\nu}=-1.$}. Hence we will be considering the lightlike trajectories with the null ray condition defined below, 
\begin{equation}
    g_{\mu\nu}\dot{x}^{\mu}\dot{x}^{\nu}=0\,.\label{2.5}
\end{equation}. Hence,
\begin{equation} \label{rev3}
    -\Big(1-\frac{2M}{r}\Big)\dot{t}^{2}+\frac{\dot{r}^{2}}{1-\frac{2M}{r}}+r^{2}\dot{\phi}^{2}=0\,.
\end{equation}
Then we write the $r$ component of equation of motion in terms of $\phi$ component. To get that, we divide (\ref{rev1}) by (\ref{rev2}) and also  define the impact parameter as 
\begin{equation} \label{impactparam}
\lambda=\frac{L}{E}.
\end{equation}
The we get, 
\begin{equation} \label{rev4}
    \frac{dt}{d\phi}=\frac{\dot{t}}{\dot{\phi}}=\frac{ r^{2}}{\lambda(1-\frac{2M}{r})}\,.
\end{equation}
The using (\ref{rev3}) we get,
\begin{equation} \label{rev5}
    \Big(\frac{dr}{d\phi}\Big)^{2}=\frac{r^{4}}{\lambda^{2}}-r^{2}\Big(1-\frac{2M}{r}\Big)\,.
\end{equation}
Equations (\ref{rev4}) and (\ref{rev5}) have all the necessary information regarding the behaviour of lightlike geodesics. Considering equation (\ref{rev5}) and taking the $\phi$ derivative of this equation we get,
\begin{equation}
    2\frac{dr}{d\phi}\frac{d^{2}r}{d\phi^{2}}=\Big(\frac{4\,r^{3}}{\lambda^{2}}-2\,r+2M\Big)\frac{dr}{d\phi}\,.
\end{equation}
For circular light-like geodesics we must have $\frac{dr}{d\phi}=0$ and $\frac{d^{2}r}{d\phi^{2}}=0$ which gives:
\begin{align}
    \begin{split}
        & \frac{r^{4}}{\lambda^{2}}-r^{2}\Big(1-\frac{2M}{r}\Big)=0\,, \\ & \frac{4\,r^{3}}{\lambda^{2}}-2r+2\,M=0\,.
    \end{split}
\end{align}
Eliminating b from the above equations we get $r=3\,M$. We have thus shown that there is a circular lightlike geodesic (or photon ring) $r=3\,M$. As we can choose any plane through the origin as our equatorial plane, there is actually a photon ring at this radius in the sense that every great circle on this sphere is a lightlike geodesic. However, the photon rings at $r=3\,M$ are unstable in
the following sense: A lightlike geodesic with an initial condition that deviates slightly
from that of a photon ring at $r=3\,M$ will spiral away from $r=3\,M$ and either go to infinity or to the horizon.
\par 
Now that we have understood how the null geodesics behave in this geometry, it's time to focus on deriving actual observable quantities that one can measure. For this, we go on to study them in the upcoming sections one by one starting with calculating deflection angles.
\subsection{Formula for deflection angle}
Let's set up the problem that we want to address. We consider a light ray that comes in from infinity and then goes through a minimum radius value at $r=r_{min}$, and then escapes back to infinity. Due to the geometry of spacetime around the central black hole, which for us is a Schwarzchild one, there will be a deflection. This deflection is simply because the rectilinear propagation of light will not be observed in this non-trivial geometry. The deflection angle measures the degree of this deviation from rectilinear propagation. In the following discussion, we will express this in terms of $r_{min}$ and the mass of the central object.
\par
We start out with (2.7) and determine the ratio $\frac{1}{\lambda^{2}}$ at $r=r_{min}$. This gives
\begin{equation}
    \frac{1}{\lambda^{2}}=\frac{1}{r^{2}_{min}}-\frac{2M}{r^{3}_{min}}
\end{equation}
Hence replacing this in (2.7) we get,
\begin{equation}
    d\phi=\frac{\pm dr}{\sqrt{(\frac{1}{r^{2}_{min}}-\frac{2M}{r_{min}^{3}})r^{4}-r^{2}+2M\,r}}
\end{equation}
which, on integrating over the coordinates of the light ray, leads to 
\begin{align}
\begin{split}
  &  \pi+\hat{\alpha}=2\int^{\infty}_{r_{min}}\frac{r_{min}dr}{\sqrt{(1-\frac{2M}{r_{min}})r^{4}-r^{2}_{min}r^{2}+2M\,r^{2}_{min}r}} \\ & =2\int^{\infty}_{r_{min}}\frac{r_{min}dr}{\sqrt{r^{2}(r^{2}-r^{2}_{min})-\frac{2M}{r_{min}}(r^{3}-r_{min}^{3})}}=2\int^{\infty}_{r_{min}}\frac{r_{min}dr}{\sqrt{1-\frac{2M}{r_{min}}\frac{(r^{3}-r_{min}^{3})}{r(r^{2}-r^{2}_{min})}}r\sqrt{r^{2}-r^{2}_{min}}}\\ & =2\int^{\infty}_{r_{min}}\Big\{1+\frac{1}{2}\frac{2M}{r_{min}}\frac{(r^{3}-r_{min}^{3})}{r(r^{2}-r^{2}_{min})}+\mathcal{O}\Big(\frac{2M}{r_{min}}\Big)^{2}\Big\}\frac{r_{min}dr}{r\sqrt{r^{2}-r^{2}_{min}}}\\ & =2\int^{\infty}_{r_{min}}\frac{r_{min}dr}{r\sqrt{r^{2}-r^{2}_{min}}}+\frac{2M}{r_{min}}\int^{\infty}_{r_{min}}\frac{(r^{3}-r^{3}_{min})r_{min}dr}{r^{2}(r^{2}-r^{2}_{min})^{3/2}}+\mathcal{O}\Big(\frac{2M}{r_{min}}\Big)^{2}=\pi+\frac{4M}{r_{min}}+\mathcal{O}\Big(\frac{2M}{r_{min}}\Big)^{2}
    \end{split}
\end{align}
Neglecting higher-order terms we get 
\begin{equation} \label{rev7}
    \hat \alpha=\frac{4\,M}{r_{min}}
\end{equation}
Let us look at a point or two about this derivation:
\begin{itemize}
\item From the derivation, it is clear that the integrand has a singularity at the lower bound $r=r_{min}$. A more detailed analysis shows that the integral is finite for all values of $r=r_{min}$ that is bigger than $\frac{3}{2}r_{s}$, where $r_{s}=2\,M$. If we consider a sequence of light rays with $r=r_{min}$ approaching $\frac{3}{2}r_{s}$ from above, the deflection angle $\delta$ becomes bigger and bigger, which
means that the light rays make more and more turns around the centre. In the limit $r_{min}\rightarrow \frac{3}{2}\,r_{s}$ the integral goes to infinity, and the limiting light ray spirals asymptotically towards a circle at $r=\frac{3}{2}\,r_{s}$. If an unstable photon ring is approached, the deflection angle goes to infinity, the singularity being logarithmic. 
\item The second point comes during the discussion on taking the Taylor series expansion in $\frac{r_{s}}{r_{min}}$ in the fourth step. The asymptotic behaviour of light rays for $r_{min}$  approaching $\frac{3}{2}r_{s}$ is relevant only for black holes and for ultracompact stars. For an ordinary star, like our Sun, $r_{min}$ takes a value much bigger than $r_{s}$. It is under this consideration that we take a Taylor series expansion in $\frac{r_{s}}{r_{min}}$.
 
\end{itemize}
\subsection{Shapiro time}
Combining equation (\ref{rev4}) and (\ref{rev5}) allows us to calculate the time taken by the light ray in this particular geometry. Let's consider a light ray that starts at a radius $r_{L}$, passes through a minimum of radius $r_{min}$ and terminates at a radius $r_{o}$. From equations (\ref{rev4}) and (\ref{rev5}) we get,
\begin{equation} \label{rev6}
    \Big( \frac{dr}{dt}\Big)^{2}=\Big( \frac{dr}{d\phi}\Big)^{2}\Big( \frac{d\phi}{dt}\Big)^{2}=\Big\{\frac{r^{4}}{\lambda^{2}}-r^{2}\Big(1-\frac{2M}{r}\Big)\Big\}\frac{(1-\frac{2M}{r})^{2}}{\lambda^{2}r^{4}}=\Big\{\frac{r^{3}}{\lambda^{2}}-r\Big(1-\frac{2M}{r}\Big)\Big\}\frac{(r-2M)^{2}}{\lambda^{2}r^{5}}\,.
\end{equation}
At, $r=r_{min}$ (\ref{rev6}) equation reads:
\begin{equation}
   \frac{r_{min}^{4}}{\lambda^{2}}-r_{min}^{2}\Big(1-\frac{2M}{r_{min}}\Big)=0\implies \frac{1}{\lambda^{2}}=\frac{r_{min}-2M}{r^{3}_{min}}\,.
\end{equation}
Finally (\ref{rev6}) becomes,
\begin{equation}
    \Big(\frac{dr}{dt}\Big)^{2}=\Big\{\frac{(r_{min}-2M)r^{3}}{r^{3}_{min}}-r\Big(1-\frac{2M}{r}\Big)\Big\}\frac{(r-2M)r^{3}_{min}}{(r_{min}-2M)r^{5}}
\end{equation}
which gives
\begin{equation}
    dt=\frac{\pm \sqrt{r_{min}-2M}\,r^{5/2}dr}{(r-2M)r^{3/2}_{min}\sqrt{\frac{(r_{min}-2M)r^{3}}{r^{3}_{min}}-r\Big(1-\frac{2M}{r}\Big)}}\,.
\end{equation}
Integrating we get the travel time for the light ray,
\begin{equation}
    \Delta t=\Big(-\int^{r_{min}}_{r_{L}}+\int^{r_{O}}_{r_{min}}\Big)\frac{ \sqrt{r_{min}-2M}r^{5/2}dr}{(r-2M)r^{3/2}_{min}\sqrt{\frac{(r_{min}-2M)r^{3}}{r^{3}_{min}}-r\Big(1-\frac{2M}{r}\Big)}}\,,
\end{equation}
where the signs on the right-hand side had to be chosen in such a way that the time
coordinate is always increasing along the light ray. One can perform this integral exactly in terms of an elliptic integral. However, $r_{min}\gg r_{s}$ we can make a Taylor approximation, in exactly the same way as we did it for the deflection formula and obtain:
\begin{align}
\begin{split}
   &  \Delta t=(\int^{r_{L}}_{r_{min}}+\int^{r_{O}}_{r_{min}})\frac{rdr}{\sqrt{r^{2}-r^{2}_{min}}}+2M(\int^{r_{L}}_{r_{min}}+\int^{r_{O}}_{r_{min}})\frac{dr}{\sqrt{r^{2}-r^{2}_{min}}}\\ & +\frac{2Mr_{min}}{2}(\int^{r_{L}}_{r_{min}}+\int^{r_{O}}_{r_{min}})\frac{dr}{\sqrt{r-r_{min}}(r+r_{min})^{3/2}}+....\,.
    \end{split}
\end{align}
The zeroth-order term is, of course, the Euclidean travel time for a light ray with speed
c along a straight line. The deviation of the general-relativistic calculation from this
zeroth-order term is known as the Shapiro time delay \cite{Will:2016sgx}.
\par
I. Shapiro suggested using this effect as the fourth test of general relativity (after perihelion precession, light deflection and gravitational redshift). In the first experiment, a strong radio signal was sent to Venus when it was in opposition to the Earth, and the time was measured until the signal arrived back on the Earth after being reflected in Venus's atmosphere. Later experiments were done with transponders on spacecraft, which sent the signal back with increased intensity. The best measurement to date was done with the Cassini spacecraft in 2002. The general-relativistic time delay was verified to be within an accuracy of 0.001$ \%$ \cite{Will:2016sgx}.

\subsection{Angular radius of Einstein rings}
Einstein ring can be observed when the light source and observer are perfectly aligned (directly opposite to each other). We want to determine the angular radius $\theta_{E}$ of the Einstein ring in dependence on the radius coordinate $r_{L}$ of the light source, the radius coordinate $r_{O}$ of the observer and the Schwarzschild radius $(=2M)$. We use the formula 
\begin{equation}
    d\phi=\frac{\pm dr}{\sqrt{\frac{(r_{min}-\frac{2M}{c^{2}})r^{4}}{r^{3}_{min}}-r^{2}+2Mr}}\,.
\end{equation}
Integrating over the light ray gives
\begin{equation}
    \pi=(\int^{r_{L}}_{r_{min}}+\int^{r_{O}}_{r_{min}})\frac{ dr}{\sqrt{\frac{(r_{min}-\frac{2M}{c^{2}})r^{4}}{r^{3}_{min}}-r^{2}+2Mr}}\,.
\end{equation}
The integration gives 
\begin{equation}
    r_{min}=f(r_{L},r_{O},2M)\,.
\end{equation}
With $r_{min}$ determined, the angular radius of the Einstein ring can be obtained by
\begin{equation}
    \tan\theta_{E}=\frac{r d\phi}{(1-\frac{2M}{r})^{-1/2}dr}|_{r=r_{O}}\,.
\end{equation}

\section{Lensing for Kerr-Newman black hole} \label{sec4}
In Sec.~(\ref{sec1}) we discussed the gravitational lensing in Schwarzschild spacetime. We expand the integrand around the turning point to get a simplified result, but in principle, we can calculate the exact deflection angle. In this section, first and foremost, we will generalize the result of lensing for rotating spacetime.   We will calculate the deflection angle for the light rays in Kerr-Newman spacetime which is an axisymmetric spacetime. One can reproduce the result for the Kerr and Schwarzschild case as special limits. We will also discuss the strong deflection angle's analytical form. We will closely follow the notation of \cite{Ghosh:2022mka}.

In Boyer-Lindquist the line element is given by  ,
\begin{eqnarray}
    ds^2&=&-\frac{\Delta}{\Sigma}(dt-a\,\sin^2\theta\,d\phi)^2+\Sigma\Big(\frac{dr^2}{\Delta}+d\theta^2\Big)+\frac{\sin^2\theta}{\Sigma}\Big(a\,dt-(r^2+a^2)\,d\phi\Big)^2 
    \label{211a}
\end{eqnarray}
where,
\begin{equation}
 \Sigma=r^2+a^2 \cos^2\theta\\,
 \Delta(r)=\Delta(r)=(r^2+a^2)-2m\,r+Q^2\,. \label{2200}
\end{equation}
In (\ref{211}) and (\ref{2200}), $m\ge0$, $Q$ and $a$ are respectively the ADM mass, charge,  and angular momentum of the black hole. As we know the particle lagrangian, specifically for the photons is given by (\ref{1}), 
with $\dot{x}^\mu=\frac{dx^\mu}{d\Tilde{\lambda}}$ for some convenient parameter $\Tilde{\lambda}$. The metric coefficients are independent of $t$ and $\phi$, implying that we will have two conserved quantities along the photon trajectory: energy $E$ and angular momentum $L$. Using the constants of motion and the null ray condition in (\ref{2.5}) the null geodesic equations can be written as,
\begin{align}
    \begin{split}
   & \rho^4 \dot{r}^2=((r^2+a^2)E-aL)^2-\Delta(r)((L-a\, E)^2+\Bar{K}):=R^2(r)\,,\\&
    \Sigma \dot{\phi}=-\Big(a\, E-\frac{L}{\sin^2\theta}\Big)+\frac{a[E(r^2+a^2)-a\, L]}{\Delta}\,,\\&
    \Sigma \dot{t}=-a\,(a\, E\,\sin^2\theta-L)+\frac{(r^2+a^2)[E(r^2+a^2)-a\, L]}{\Delta}\,,\\&
    \rho^4 \dot{\theta}^2=\Bar{K}+\cos^2\theta\Big(a^2\, E^2-\frac{L^2}{\sin^2\theta}\Big):=\Theta(\theta)^2\,.\label{211}
\end{split}
\end{align}
where the Carter constant, which results from the separability of the Hamilton-Jacobi equation, is $\Bar{K}$. For our purpose, we will look  upon into the equatorial case only i.e, $\theta=\frac{\pi}{2},\dot{\theta}=0$ which automatically implies $\Bar{K}=0$, and then will investigate the non-equatorial case also. We can use the geodesic equations to calculate the equatorial deflection angle. Rather in the next subsection we will provide the deflection angle calculations and reproduce the result for Kerr by taking the limit $Q=0$.
\subsection{Deflection angle for Kerr-Newman spacetime: }
From (\ref{211}) we can write the radial geodesic equations as follows,
\begin{eqnarray}
   \frac{\dot{r}^2}{L^2}+V(r)=\frac{1}{\lambda^2}\,,\label{212}
\end{eqnarray}
with the effective potential,
\begin{eqnarray}
V(r)=\frac{1}{r^2}\Big[1-\frac{a^2}{\lambda^2}+\Big(1-\frac{a}{\lambda}\Big)^2\Big(-\frac{2m}{r}+\frac{Q^2}{r^2}\Big)\Big]
\label{213}
\end{eqnarray}
and $\lambda$ is the impact parameter defined in (\ref{impactparam}).\par
Now, think of light rays that originate at infinity, pass through the black hole, and then return to infinity to reach the observer. The closest approach to the black hole, $r_0$, will be the radial turning point for these light rays, determined by,
\begin{eqnarray} \label{eqq1}
   \Big(\frac{\dot{r}^2}{L^2}\Big)\Big |_{r=r_0}=\frac{1}{\lambda^2}-V(r_0)=0.\label{220}
\end{eqnarray}
From (\ref{eqq1}) we get,
\begin{eqnarray}
     {r_0}^4-\lambda^2\Big(1-\frac{a^2}{\lambda^2}\Big){r_0}^2+2\,m\lambda^2\Big(1-\frac{a}{\lambda}\Big)^2 {r_0}=Q^2\lambda^2\Big(1-\frac{a}{\lambda}\Big)^2 \label{221}
\end{eqnarray}
Solving, (\ref{221}) we get,
\begin{align}
    \begin{split}
   &  r_0(\lambda)=\frac{\lambda}{\sqrt{6}}\sqrt{1-\omega^2}\Big[\sqrt{1+\hat \gamma}+\sqrt{2-\hat \gamma-\frac{3\sqrt{6}m(1-\omega)^2}{\lambda(1-\omega^2)^{3/2}\sqrt{1+\hat \gamma}}}\Big]\label{222}
     \end{split}
\end{align}
where $\omega,\chi,\rho,\hat{\gamma}$ are given by,
\begin{align}
    \begin{split}
  & \omega=\frac{a}{\lambda},\,\, \rho=\frac{12Q^2}{\lambda^2(1+\omega)^2}\,\,,\hat{\gamma} =\sqrt{1-\rho}\cos\Big(\frac{2\chi}{3}\Big), \\&
     \chi=\arccos\Big(\frac{3\sqrt{3}\,m(1-\omega)^2}{\lambda(1-\omega^2)^{3/2}(1-\rho)^{3/4}}\sqrt{1-\frac{\lambda^2(1+\omega)^3}{54\,m^2(1-\omega)}\Big[1+3\rho-(1-\rho)^{3/2}\Big]}\Big)\,.\label{223}
     \end{split}
\end{align}
One can convince that the non-zero charge of the black holes gives a repulsive effect on the light rays but the effect of spin attracts the light rays toward the black hole.
\subsection{Photon sphere radius and critical impact parameter}
In this subsection, we will define the photon sphere and the critical impact parameter which will be useful in the subsequent sections. The radius of the photon sphere is the value of $r$ where the effective potential attains its maximum value. 
\begin{eqnarray}
   \frac{\partial V(r)}{\partial r}\Big|_{r=r_c}=0\,. \label{vmax}
\end{eqnarray}
From (\ref{vmax}) we can find out 
\begin{eqnarray}
     r_c(\lambda_c)={\frac{3\,m\,\hat \zeta}{2}\Big[1+\sqrt{1-\zeta}\Big]}\label{33}
\end{eqnarray}
 where, $$\quad \hat \zeta= \Big(\frac{1-\frac{a}{\lambda_c}}{1+\frac{a}{\lambda_c}}\Big), \quad \zeta=\frac{8\,Q^2}{9\,m^2}\hat \zeta\,. $$\par 
 Note that the turning point of the photon is $r_0$ defined in (\ref{222}) and attains its minimum value at $r_c$ with the corresponding impact parameter $\lambda_c.$ If for some $\lambda,$ $r_0$ becomes less than $r_c$ then the photon will fall into the black hole. $\lambda_c$ is called the critical impact parameter. Finally, substituting equation (\ref{222}) into (\ref{33}) we can write  $\lambda_c=\lambda_c(a,Q,l)\,\, \textrm{and} \,\, r_c=r_c(a,Q,l).$\par
\subsection{An exact analytical computation of the deflection angle}\label{3.3}
In this subsection, we will compute the exact photon deflection angle near a Kerr-Newman black hole. To do this we can choose any polar plane but to keep things simple we choose the equatorial plane $\theta=\frac{\pi}{2}$ and $\Dot{\theta}=0$. Furthermore to investigate the observational signature we need to employ the strong deflection limit. To do this we have to evaluate the deflection angle in the mentioned plane and take the strong limit. We can rather do the reverse procedure also, that is from the beginning take the strong limit and then calculate the deflection angle at this limit. The second one is simpler to do. We adopt this track while discussing the observational signature.  \par 
 The analysis has been done in\cite{Iyer:2009wa, Hsiao:2019ohy}. We review the analysis here.
Before proceeding with the computation we define the following coordinate, 
\begin{eqnarray}
   u:=\frac{1}{r}\,.
\end{eqnarray}
Then combining the first two equations in (\ref{211}) we get,
\begin{eqnarray}
   \Big(\frac{du}{d\phi}\Big)^2=\Big(\frac{du}{dr}\frac{dr}{d\phi}\Big)^2=\frac{r^2}{r^6}\frac{\dot{r}^2}{\dot{\phi}^2}=(u^4)\frac{\dot{r}^2}{\dot{\phi}^2}\,.
\label{42}\end{eqnarray}
 Now our goal is to rewrite the $\dot{r}$ and $\dot{\phi}$ as function of $u$.
We have,
\begin{eqnarray}
    \dot{r}^2&=&L^2\Big(\frac{1}{\lambda^2}-V(r)\Big)\nonumber\\&=&L^2\Bigg[\frac{1}{\lambda^2}-\Big[\frac{1}{r^2}\Big(1-\frac{a^2}{\lambda^2}+(1-\frac{a}{\lambda})^2(-\frac{2m}{r}+\frac{Q^2}{r^2}\Big)\Big]\Bigg]\nonumber\\&=&L^2\Big[\frac{1}{\lambda^2}-u^2\Big(1-\frac{a^2}{\lambda^2}\big)-Q^2\Big(1-\frac{a}{\lambda}\Big)^2 u^4+2m\Big(1-\frac{a}{\lambda}\Big)^2 u^3\Big]:=L^2 B(u)\,.\label{43}
\end{eqnarray}
Combining equation (\ref{43}) and (\ref{211}) and using the equation (\ref{42}) we  get,
\begin{eqnarray}
   \Big(\frac{du}{d\phi}\Big)^2=\Bigg[\frac{1-2mu+(a^2+Q^2)u^2}{1-(2mu-Q^2u^2)(1-\frac{a}{\lambda})}\Bigg]^2 B(u)\,.\label{45}
\end{eqnarray} 
The photon deflection angle $\hat{\alpha}$ can be calculated by integrating the equation (\ref{45}) over $u$ from $0$ to $\frac{1}{\sqrt{r}}$, where $r_0$ is the turning point, and then evaluating the resulting expression at the critical value $r_c$. \cite{Bozza:2002af}, 
\begin{eqnarray}
   \hat{\alpha}=-\pi+2\int_0 ^\frac{1}{r_0}\,du\,\Bigg[\frac{1-(2mu-Q^2u^2)(1-\frac{a}{\lambda})}{1-2mu+(a^2+Q^2)u^2}\Bigg]\frac{1}{\sqrt{B(u)}}\,\label{46s}
\end{eqnarray}
with $\omega=\frac{a}{\lambda}.$ This integral can be computed exactly. We give the final result here. Details of the computation of this integral are given in the Appendix~\ref{appendixA}.
\begin{align}
    \begin{split}
       \hat\alpha & = -\pi+\frac{4}{(1-\omega)\sqrt{Q^2(u_4-u_2)(u_3-u_1)}}  \\ &
        \Bigg[\frac{G_{+}+K_{Q_{+}}u_1}{u_{+}-u_1}[\Pi(n_+,k)-\Pi(n_+,\phi,k)]+\frac{G_{-}+K_{Q_{-}}u_1}{u_--u_1}[\Pi(n_+,k)-\Pi(n_+,\phi,k)]\\ &
        - \frac{G_{+}+K_{Q_{+}}u_4}{u_{+}-u_4}[\Pi(n_+,k)-\Pi[n_+,\phi,k]-F(\frac{\pi}{2},k)+F(\phi,k)]\\ &
        -\frac{G_{-}+K_{Q_{-}}u_4}{u_{-}-u_4}[\Pi(n_-,k)-\Pi[n_-,\phi,k]-F(\frac{\pi}{2},k)+F(\phi,k)]\Bigg]
    \end{split}
\end{align}
where,
\begin{eqnarray}
u_1&=&\frac{X_1-2m-X_2}{4m{r_0}}\,, u_2=\frac{1}{{r_0}}\,, u_3=\frac{X_1-2m+X_2}{4m r_0}\,,u_4=\frac{2m}{Q^2}-\frac{X_1}{2m r_0}\,,\nonumber\\
    n_{\pm}&=&\frac{u_2-u_1}{u_{\pm}-u_1}\Big[1+\frac{2mQ^2(1-r_0 u_{\pm})}{4m^2r_0-Q^2(X_1+2m)}\Big]\,,
    k^2=\frac{(X_2+6m-X_1)[8m^2r_0-Q^2(X_2-2m+3X_1)]}{4X_2[4m^2r_0-Q^2(X_2+2m)]},\nonumber\\
    \psi_{0}&=& \arcsin \sqrt{\frac{(X_2+2m-X_1)[4m^2r_0-Q^2(X_2+2m)]}{(X_2+6m-X_1)(4m^2r_0-Q^2 X_1)}},\nonumber\\
    G_+&=&\frac{2m(1-\omega)\Big(m+\sqrt{m^2-(a^2+Q^2)}\Big)-(a^2+Q^2)}{2(a^2+Q^2)\sqrt{m^2-(a^2+Q^2)}},\nonumber\\
    G_-&=&\frac{(a^2+Q^2)-2m(1-\omega)\Big(m-\sqrt{m^2-(a^2+Q^2)\Big)}}{2(a^2+Q^2)\sqrt{m^2-(a^2+Q^2)}},\nonumber\\
     G_{Q+}&=&\frac{-Q^2(1-\omega)\Big(m+\sqrt{m^2-(a^2+Q^2)}\Big)}{2(a^2+Q^2)\sqrt{m^2-(a^2+Q^2)}}, G_{Q-}= \frac{Q^2(1-\omega)\Big(m-\sqrt{m^2-(a^2+Q^2)}\Big)}{2(a^2+Q^2)\sqrt{m^2-(a^2+Q^2)}}. \label{neweqs}
\end{eqnarray}
Note that, $\Pi(n_+,\psi_{0},k)$ and $\Pi(n_+,k)$ are the incomplete and complete elliptic integral of the third kind respectively. Also, $F(\psi_{0},k)$ and $F(\frac{\pi}{2},k)$ are the incomplete and complete elliptic integral of the first kind. Also, 
\begin{align}
\begin{split}
& \textstyle{X_1=\frac{2m(Q^2+4mr_0)}{3Q^2}
+\frac{8m^2 r_0}{3Q^2}\sqrt{1+\frac{Q^2}{2m^2}\Big(\frac{m}{r_0}-\frac{3(1+\omega)}{2(1-\omega)}-\frac{Q^2}{r_0^2}}\Big)\,      \cos\Big(\frac{\delta}{3}+\frac{2\pi}{3}\Big)}\,.\label{equ21s}
\end{split}
\end{align}
where,
\begin{align}
\begin{split}
   & \textstyle{\delta\,\, =\,\, \arccos\Big(\frac{-8m^3 r_0^3-3\,m\,Q^2 r_0^2\Big(2m-\frac{3r_0(1+\omega)}{(1-\omega)}\Big)-3\,Q^4 r_0\Big(5m-\frac{3r_0(1+\omega)}{(1-\omega)}\Big)+10\,Q^6}{\Big[4m^2 r_0^2+Q^2 r_0\Big(2m-\frac{3r_0(1+\omega)}{(1-\omega)}\Big)-2\,Q^4\Big]^{3/2}}\Big)}\,\label{equ21}
\end{split}
\end{align}
and $X_2$ can be obtained from the following equations after inserting $X_1$ from (\ref{equ21}),
\begin{equation}
 \Big[X_2^2-(X_1^2-2m)^2\Big]\Bigg(\frac{1}{8m r_0^3}-\frac{Q^2X_1}{32m^3 r_0^4}\Bigg)=\frac{1}{\lambda^2(1-\omega)^2}\,.
     \end{equation}

So, the exact deflection angle can be derived as discussed. One can reproduce the results for the Schwarzschild black hole in the limit $a\rightarrow 0$ and $Q\rightarrow 0$ and Kerr black hole in the limit $a\rightarrow 0$.

\subsection{Strong deflection analysis for axisymmetric spacetime}\label{3.4}
In this subsection, we inspect the strong limit of the equatorial deflection angle ($\theta=\frac{\pi}{2}$) of the light rays discussed in the previous section in order to gain further understanding of the deflection angle for our context and make touch with the feasible observational signature. The strong deflection limit of (\ref{431}) is challenging to take directly. It will be simpler to perform the integration after taking the strong field limit of the integrand of (\ref{46}). We shall only take into account the photons having the turning point very close to the photon sphere's radius, as was previously specified.\par
The metric on the equatorial plane has the following structure:
\begin{eqnarray}
   ds^2=-\Bar{A}(r)dt^2+\Bar{B}(r)dr^2+\Bar{C}(r)d\phi^2-\Bar{D}(r)dt\,d\phi
\end{eqnarray}
with
\begin{align}\begin{split}
   &\Bar{A}(r)=\frac{(\Delta(r)-a^2)}{\Sigma}\,,\,\,\, \Bar{B}(r)=\frac{\Sigma}{\Delta}\,,
   \Bar{C}(r)=\frac{1}{\Sigma}\Big[(r^2+a^2+l^2)^2-\Delta(r)a^2\Big]\,,\\&
   \Bar{D}(r)=\frac{2}{\Sigma}\Big[(r^2+a^2+l^2)a-\Delta(r)a\Big]\,.
\end{split}\end{align}
It should be noted that all metric components are evaluated in the equatorial plane. We have already seen that the spacetime admits two conserved quantities $E$ and $L$ due to the existing symmetries. To keep things simple, we set $E=1$. As a result, the impact parameter $\lambda=L$. Using the fact that at the distance of closest approach $r=r_0$ and $\dot r=0$, we get the following from (\ref{212}),
\begin{align}
\begin{split}
  & L=\frac{-\Bar{D}_0+\sqrt{4\Bar{A}_0\Bar{C}_0+\Bar{D}_0^2}}{2\Bar{A}_0}\\ &
  =\frac{r_0 \left(r_0^2 \sqrt{a^2-2\,m\, r_0+Q^2+r_0^2}+a \left(Q^2-2\,m\, r_0\right)\right)}{Q^2 r_0+l^2 \left(r_0-2\,m\,\right)+r_0^2 \left(r_0-2\,m\,\right)}\,.
\end{split} \label{433}
\end{align}
The subscript $``0"$ denotes functions are evaluated at $r=r_0$. From the equation of motion of $\phi$ (the second equation of (\ref{211})) we get,
\begin{eqnarray}\label{az}
   \phi(r_0)=2\int_{r_{0}}^\infty\frac{\sqrt{\Bar{B}\Bar{A}_0}(\Bar{D}+2L\Bar{A})}{\sqrt{4\Bar{AC}+\Bar{D}^2}\sqrt{\Bar{CA}_0-\Bar{AC}_0+L(\Bar{AD}_0-\Bar{DA}_0)}}\,dr\,.
\end{eqnarray}
In the strong limit, we only consider the photons having closest approach $r_c$ near to the radius of the photon sphere. To implement the limit, one can expand the deflection angle $\hat\alpha$ around $r_c$ or $\lambda_c$ and do the radial integral. When the turning point $r_0$ is greater than the radius of the photon sphere $r_c$ then we get a finite deflection angle otherwise the photon will be caught by the black hole and the deflection angle diverges. Following the method developed here \cite{Bozza:2002zj, Bozza:2002af}, it is easy to find out the nature of the divergence in the deflection angle when the photons are very close to the radius of the photon sphere $r=r_c$. One can define two variables $y,z$ as,
\begin{eqnarray}
   z_1=\Bar{A}(r),
   z_2=\frac{z_1-z_{1,0}}{1-z_{1,0}}\,.
\end{eqnarray}
Now, the azimuthal angle defined in (\ref{az}) can be expressed in terms of these two new variables,
\begin{eqnarray}
   \phi(r_0)=\int_0^1\Bar{R}(z_2,r_0)\Bar{F}(z_2,r_0) dz_2\label{59}
\end{eqnarray}
where,
\begin{eqnarray}
   \Bar{R}(z_2,r_0)&=&\frac{2(1-z_{1,0})}{A'}\frac{\sqrt{\Bar{B}\Bar{A}_0}(\Bar{D}+2L\Bar{A})}{\sqrt{4\Bar{AC}^2+\Bar{CD}^2}}\\
   \Bar{F}(z_2,r_0)&=&\frac{1}{\sqrt{\frac{1}{\Bar{C}}(\Bar{CA}_0-\Bar{AC}_0+L(\Bar{AD}_0-\Bar{DA}_0)})}=\frac{1}{\sqrt{\Bar{H}}}\,,\label{H1}\\
   \Bar{H} & =&\frac{1}{\Bar{C}}(\Bar{CA}_0-\Bar{AC}_0+L(\Bar{AD}_0-\Bar{DA}_0))\,.\label{H}
\end{eqnarray}
The function $\Bar{R}(z,r 0)$ is regular for any $z$ and $r_0$ values, whereas the function $\Bar{F}(z,r 0)$ is divergent for $z=0$, i.e. at $r=r_0$. As a result, after extracting the divergent part, one can rewrite (\ref{az}).
\begin{eqnarray}
   \phi(r_0)=\phi_{\Bar{R}}(r_0)+\phi_{\Bar{F}}(r_0)
\end{eqnarray}
where the divergent part can be written as,
\begin{eqnarray}
   \phi_{\Bar{F}}(r_0)=\int_0^1\Bar{R}(z_2=0,r_c)\Bar{F}_0(z_2,r_0)dz\,.\label{513}
\end{eqnarray}
We know that the deflection angle should diverge at $r_0=r_c$, indicating that the photon has been captured by the black hole. Next, we want to determine the nature of the divergence. Examining the denominator will allow us to determine the nature of the divergence (\ref{59}). In order to achieve this, we Taylor expand the denominator of $\Bar{F}_0(r_0,z_2)$(\ref{514}) around $z_2=0$.
\begin{eqnarray}
   \Bar{F}_{0}(z_2,r_0)\approx \frac{1}{\sqrt{\sigma_1(r_0)z+\sigma_2(r_0) z_{2}^2+\mathcal{O}(z^3})}\,.\label{514}
\end{eqnarray}
It is worth noting that if $\sigma_1(r_0)=0,$ (this occurs when $r_0$ coincides with the radius of the photon sphere, \cite{Claudel:2000yi}), then it is clear from (\ref{514}) that the leading term is $\frac{1}{z}$ in the small $z$ limit. As a result, after integration, we get a logarithmic divergence, as shown in (\ref{523}).
To find $\sigma_1$ and $\sigma_2$, first Taylor expand $\Bar{H}$, which is defined in (\ref{H}).
\begin{eqnarray}
   \Bar{H}(z,r_0)=\Bar{H}(0,r_0)+\frac{\partial\Bar{H}}{\partial z_2}\Big|_{z_2=0}z_2+\frac{1}{2!}\frac{\partial^2\Bar{H}}{\partial z_2^2}\Big|_{z_2=0}z_2^2+\mathcal{O}(z_2^3),\text{with}\,\,\Bar{H}(0,r_0)=0\,.
\end{eqnarray}
Therefore using (\ref{H1}) we can identify $\sigma_1$ and $\sigma_2$ as ,
\begin{eqnarray}
   \sigma_1&:=&\frac{\partial \Bar{H}}{\partial z_2}\Big|_{z_2=0}\nonumber\\
   &=&\frac{1-\Bar{A}_0}{\Bar{A}_0'\Bar{C}_0}\Bigg(\Bar{A}_0\Bar{C}_0'-\Bar{A}_0'\Bar{C}_0-L\Big(\Bar{A}_0\Bar{D}_0'-\Bar{A}_0'\Bar{D}_0\Big)\Bigg)
\end{eqnarray}
and
\begin{align}
\begin{split}
  &  \sigma_2:=\frac{1}{2!}\frac{\partial^2 \Bar{H}}{\partial z_2^2}\Big|_{z_2=0}\\ &
   =\frac{(1-\Bar{A}_0)^2}{2\Bar{C}_0^2\Bar{A}_0'^3}\Big[2\Bar{C}_0\Bar{C}_0'\Bar{A}_0'^2+\Big(\Bar{C}_0\Bar{C}_0''-2\Bar{C}_0'^2\Big)\Bar{A}_0\Bar{A}_0'-\Bar{C}_0\Bar{C}_0'\Bar{A}_0\Bar{A}_0''  \\ &
   +L\Big(\Bar{A}_0\Bar{C}_0(\Bar{A}_0''\Bar{D}_0'-\Bar{A}_0'\Bar{D}_0'')+2\Bar{A}_0'\Bar{C}_0'(\Bar{A}_0\Bar{D}_0'-\Bar{A}_0'\Bar{D}_0)\Big)\Big]\,.
\end{split}
\end{align}
We can write the regular part as,
\begin{eqnarray}
   \phi_{\Bar{R}}(r_0)=\int_0^1\Bar{G}(z_2,r_0)dz_2\label{518}
\end{eqnarray}
where $\Bar{G}(z_2,r_0)=\Bar{R}(z_2,r_0)\Bar{F}(z_2,r_0)-\Bar{R}(z_2=0,r_c)\Bar{F}_0(z_2,r_0)$.
As we talked about the coefficient $\sigma_1=0$, in the strong deflection limit. This implies, 
\begin{eqnarray}
  \Bar{A}_0\Bar{C}_0'-\Bar{A}_0'\Bar{C}_0-L\, (\Bar{A}_0\Bar{D}_0'-\Bar{A}_0'\Bar{D}_0)\Big|_{r_0=r_c}=0\,.\label{519}
\end{eqnarray}
From (\ref{519}) we get,
\begin{align}
    \begin{split}
 &
 -2 a \left(Q^2-m r_c\right) \sqrt{a^2-2 m r_c+Q^2+r_c^2}+2 a^2 \left(Q^2-m r_c\right)+r_c^2 \left(6 m^2-5 m r_c+r_c^2\right)\\ &+Q^2 r_c (3 r_c-7 m)+2 Q^4=0
\,. \label{r0}
    \end{split}
\end{align}
It is simple to verify that the expression for the critical impact parameter $\lambda_c$ found in (\ref{433})  yields the identical expression for the radius of the photon sphere found in (\ref{33}). The photon sphere is defined yet again by (\ref{r0}), according to this. Readers who are interested in learning more about the geometry of photon spheres and several complementary definitions of photon spheres are directed to \cite{Claudel:2000yi}. \par
For fixed values of $Q,a$, we can compute the radius of the photon sphere in Kerr-Newman spacetime. In the next section, we will discuss the more general case which is the Kerr-Newman black bounce spacetime and the solutions of the photon sphere equation will be discussed there in a more general setting.\par
We may now assess the divergent integral (\ref{513}) with the help of the following.
\begin{eqnarray}
   \phi_{\Bar{F}}(r_0\approx r_c)&=&\Bar{R}(z_2=0,r_c)\int_{0}^1\frac{1}{\sqrt{\sigma_1z_2+\sigma_2z_2^2}}\,dz_2\nonumber\\
   &=&\Bar{R}(z_2=0,r_0\approx r_c)\frac{2}{\sqrt{\sigma_2}}\log\Big(\frac{\sqrt{\sigma_2}+\sqrt{\sigma_2+\sigma_1}}{\sqrt{\sigma_1}}\Big)\,.\label{521}
\end{eqnarray}
We are aware that the function $\phi_{\Bar{F}}(r_0)$ diverges when $r_0=r_c$, or when the coefficient $\sigma_1=0$. In order to obtain the nature of $\phi_{\Bar{F}}(r_0 \approx r_c)$, the idea is to expand the $\sigma_1(r_0)$ around $r_0=r_c$ up to first order and put it into (\ref{521}).
\begin{eqnarray}
   \sigma_1(r_0)&=&\frac{\partial \sigma_1}{\partial r_0}\Bigg|_{r_0=r_c}(r_0-r_c)+\mathcal{O}(r_0-r_c)^2\,, \nonumber\\
   \sigma_2(r_0)&=&\sigma_2(r_c)+\frac{\partial \sigma_2}{\partial r_0}\Bigg|_{r_0=r_c}(r_0-r_c)+\mathcal{O}(r_0-r_c)^2\,. \label{522}
\end{eqnarray}
Substituting (\ref{522}) into (\ref{521}) and using the condition (\ref{519}) we get,
\begin{eqnarray}
   \phi_{\Bar{F}}(r_0\approx r_c)&=&-\tilde{a}\log\Big(\frac{r_0}{r_c}-1\Big)+\tilde{b}+\mathcal{O}(r_0-r_c)\,.\label{523}
\end{eqnarray}
Alternatively, the equation (\ref{523}) can be written in terms of impact parameter as\cite{Bozza:2002af},
\begin{align}
\begin{split} 
  & \hat{\alpha}(\lambda)=-\Bar{a}\log\Big(\frac{\lambda}{\lambda_c}-1\Big)+\Bar{b}+\mathcal{O}(\lambda-\lambda_c) \label{sdeflection},
\end{split}
\end{align}
where the coefficients are,
\begin{eqnarray}
  \Bar{a}&=&\sqrt{\frac{2\Bar{A}_c\Bar{B}_c}{\Bar{A}_c\Bar{C}_c''-\Bar{A}_c''\Bar{C}_c+\lambda_c(\Bar{A}_c''\Bar{D}_c-\Bar{A}_c\Bar{D}_c'')}}\,,\\
  \Bar{b}&=&-\pi+b_R+\Bar{a}\log\Big(\frac{4\,\sigma_{2c}\,\Bar{C}_c}{\lambda_c\,\Bar{A}_c(\Bar{D}_c+2\lambda_c\Bar{A}_c)}\Big)\,,\\
  \lambda_c&=&L_c.
\end{eqnarray}
 The next section will show how the deflection angle (\ref{sdeflection}) varies with respect to the impact parameter. The black hole can be viewed as a lens, with its gravitational field curving the path of photons. Let $\theta=\frac{\lambda}{\boldsymbol{D_{OL}}}$ be the angular separation between the image and lens, and $\boldsymbol{D_{0l}}$ be the distance between the lens and the observer. The deflection angle (\ref{sdeflection}) can then be expressed as follows:
\begin{align}
\begin{split}
  & \hat{\alpha}({\theta})=-\Bar{a}\log(\frac{{\theta}\boldsymbol{D_{ol}}}{\lambda_c}-1)+\Bar{b}+\mathcal{O}(\lambda-\lambda_c)
  \label{426}
\end{split}
\end{align}
where $b_R=\phi_{\Bar{R}}(r_c)=\int_0^1\Bar{G}(z,r_c)dz.$ \par 
In order to do the integral we can expand $\Bar{R}(z,r_c)$  around $ z=0$ and then putting it in (\ref{513}). Formally we get the following expression,
\begin{eqnarray}
   b_R=\frac{1}{\sqrt{\sigma_2}}\int_0^1\Big(\frac{\Bar{R}(z)}{z}-\frac{\Bar{R}(0)}{z}\Big)dz\Bigg|_{r_c}=\frac{1}{\sqrt{\sigma_2}}\sum_{n=1}^\infty\frac{1}{n}\frac{\partial^n \Bar{R}}{\partial z^n}\Bigg|_{z=0}\xrightarrow[]{}\text{ finite}\,.
\end{eqnarray}
We will have multiple images of the source if the deflection angle is greater than $2\pi$. The angular radius of the Einstein ring, which is created due to the symmetric lensing of light rays coming from some distant source, can be calculated using the strong field deflection angle formula given in (\ref{426}) and the lens equation \cite{PhysRevD.77.124042,Virbhadra:2008ws}. The observational signatures of deflection angle will be discussed in the following section. 
\subsection{Non-equatorial Lensing}\label{sec3.5}
So far, we have discussed equatorial lensing. In this section we will briefly discuss about the non-equatorial lensing in Kerr-Newmann spacetime for small inclination. We assume that the inclination is $\theta=\frac{\pi}{2}-\psi$, with $\psi$ being very small. To conduct the analysis, we will closely follow the analysis presented in \cite{Bozza:2002af}.
\par 
For non-equatorial plane the carter constant $\Bar{K}\neq 0$. Rather for the small inward inclination angle, we can write down the constants in terms of inclination angle $\psi$ as follows,
\begin{eqnarray}
 L&\approx& \lambda\\
 \Bar{K}&\approx&h^2+(\lambda^2-a^2) \phi^2\,\,, \text{with}\,\, \phi\approx \frac{h}{\lambda}
\end{eqnarray}
 In principle one can parameterize the light ray coming from infinity by three parameters ($\phi, h, \lambda$). If there is no gravitational field the projection of the photon line on the equatorial plane has a minimum distance from the origin which is $\lambda$. Now for given $\lambda$, the vertical distance of the light ray from the plane is $h$ and finally $\phi$ is the inclination angle formed by the light ray with the equatorial plane.\par
Now using the $\theta$ and $\phi$ geodesic equation in (\ref{211}) and requiring $\psi$ to be small we will have,
\begin{align}
    \begin{split}
        \frac{d\psi}{d\phi}= \omega(r(\phi))\sqrt{\hat{\psi}^2-\psi^2} \,\,,\text{with}\,\,\hat{\psi}=\sqrt{\frac{h^2}{\lambda^2-a^2}+\phi^2}
    \end{split}
\end{align}
We are interested in computing the deflection angle. For that we first write down the following.
\begin{eqnarray}
 \bar{\phi_f}=\int_0^{\phi_f}d\phi\,\, \omega(\phi)
\end{eqnarray}
where $\phi_f$ is the total azimuthal shift. Then the deflection angle can be written as
\begin{eqnarray}
 {\hat \alpha}=-\pi+2\,\int_{r_0}^\infty dr \, \omega(r) \frac{d\phi}{dr}=\int_0^1\,dz\,\,\omega(r(z)) \Bar{R}(z,r_0) \Bar{F}(z,r_0)
\end{eqnarray}
where,
\begin{eqnarray}
 \omega(r)= \bar{\lambda}\,\frac{a^2+r(r-2)}{r\Big(2\,a+\lambda(r-2)\Big)+Q^2(\lambda-a)}\label{3.54}
\end{eqnarray}
with, $\bar{\lambda}=\sqrt{\lambda^2-a^2}$.
We can follow the same method as mentioned in Sec.~(\ref{3.4}) to extract the divergent part as well as the finite part  of the deflection angle as the function $\omega(r).$ They are given by,
\begin{eqnarray}
 \hat\alpha=-\hat{a}\log\Big(\frac{\lambda}{\lambda_c}-1\Big)+\hat{b},
\end{eqnarray}
where,
\begin{eqnarray}
 \hat{a}&=&\frac{\omega(z=0, r_c)\Bar{R}(z=0,r_c)}{2\sqrt{\sigma_{2c}}},\\
 \hat{b}&=&-\pi+\hat{b}_R+\hat{a}\log\Big(\frac{4\,\sigma_{2c}\,\Bar{C}_c}{\lambda_c\,\Bar{A}_c(\Bar{D}_c+2\lambda_c\Bar{A}_c)}\Big)\,,
 \end{eqnarray}
and
\begin{eqnarray}
 \hat{b}_R=\int_0^1\,dz\,[\omega(z,r_c)\Bar{R}(z,r_c)\Bar{F}(z,r_c)-\omega(z=0,r_c)\Bar{R}(z=0,r_c)\Bar{F}(z,r_c)].
\end{eqnarray}
Our interest is to find the position of the caustics, where the magnification diverges, and is given in \cite{Bozza:2002af},
$\bar{\gamma_k}=-\bar{b}+\bar{a}(\hat{b}-k\,\pi)$. Here $k$ is a positive integer. We have one caustic for the direct photon and one for the retrograde photon for each value of $k$. We will find the caustic point in the weak field limit for $k=1$ and the strong field limit for $k\ge 1$, which is the regime of interest to us. In the next section, we give the results for caustic points in a more generic setup, i.e. for the black bounce metric. By taking the limit, $l=0$, one can reproduce the results for the Kerr-Newman case. Furthermore from equation (\ref{3.54}) one can see $\omega(r)=1$ implying that for Schwarzschild metric there is no such difference between equatorial and non-equatorial case.
 \section{{Lensing for Kerr-Newman black-bounce}}\label{sec5}
Before proceeding further, we will discuss how to calculate the exact deflection angle for Kerr-Newman black-bounce spacetime \cite{Franzin:2021vnj} and then go to the observational signatures gradually. It is interesting to study because it has one more parameter (apart from mass, rotation and charge) that regularizes the central singularity. One can reproduce the results for Schwarzschild, Kerr, and Kerr-Newman spacetimes by taking appropriate limits. We start by applying the general formalism of lensing in this special kind of axisymmetric non-singular spacetime.  We will closely follow the notation of \cite{Ghosh:2022mka} throughout this section.
 \subsection{Brief review of Kerr-Newman black-bounce spacetime}
  We will  begin with a brief discussion of the null geodesics in Kerr-Newman black-bounce spacetime. To do that, we first write the corresponding metric in Boyer-Lindquist coordinate \cite{Franzin:2021vnj}, 
\begin{eqnarray}
    ds^2&=&g_{{\mu}{\nu}} dx^{\mu}dx^{\nu}\nonumber\\&=&-\frac{\Delta}{\Sigma}(a\, \sin^2 \theta\, d\phi-dt)^2 + \frac{\sin^2\theta}{\Sigma}((r^2+a^2+l^2)d\phi -a\, dt)^2+ \Sigma (\frac{dr^2}{\Delta}+d\theta^2)\label{21}
\end{eqnarray}
where
\begin{equation}
 \Sigma=r^2+l^2+a^2 \cos^2\theta\\,
 \Delta(r)=(r^2+a^2+l^2)-2m\sqrt{r^2+l^2}+Q^2\,. \label{22}
 \end{equation}
Here, $m\ge0$ is the ADM mass, $Q$ is the black hole charge parameter, and $a=\frac{J}{m}$ corresponds to the angular momentum per mass. The non-vanishing regularising parameter $l>0$ accounts for the absence of the central singularity. Keeping in mind that the radial coordinate's range in this instance is $-\infty<r<\infty$. In the $Q=0,l=0$ limit, we recover the Kerr metric. The location of the event horizon can be determined by equating $\Delta(r)=0.$
\begin{eqnarray}
     R_H&=&\sqrt{\Big[\Big(m+\sqrt{m^2-(a^2+Q^2)}\Big)^2-l^2\Big]}
\end{eqnarray}
We also need to impose the reality condition. That gives,  $$m^2-(a^2+Q^2)>0\quad \textrm{and}\quad m+\sqrt{m^2-(a^2+Q^2)}>l.$$
Following the procedure mentioned in Sec.~(\ref{sec4}) one can obtain the turning point and radius of the photon sphere. 
\subsection{Perturbative computation of the deflection angle: analytical results}\label{Sec4.2}
Following the analysis mentioned in Sec~(\ref{3.3})one can write down the integral form of the deflection angle which is given by,
\begin{eqnarray}
   \hat{\alpha}=-\pi+2\int_0 ^\frac{1}{\sqrt{r_0^2+l^2}}\frac{1}{\sqrt{1-l^2u^2}}\Bigg[\frac{1-(2mu-Q^2u^2)(1-\frac{a}{\lambda})}{1-2mu+(a^2+Q^2)u^2}\Bigg]\frac{1}{\sqrt{B(u)}}\,.\label{46}
\end{eqnarray}
The polynomial $B(u)(1-l^2u^2)$ in (\ref{46}) has \textit{degree six}. As a result, we cannot write this integral as an elliptic integral in its entirety. However, in order to make some analytical headway, we will make the following assumption: 
    \begin{equation}l^2u^2<<1.
    \end{equation}
    Then we can Taylor expand $$\frac{1}{\sqrt{1-l^2u^2}}=1+\frac{l^2u^2}{2}+\mathcal{O}(l^4u^4)$$. Finally keeping terms upto $\mathcal{O}(l^4)$ in (\ref{46}) we get,
    \begin{eqnarray}
   \hat{\alpha}&=&-\pi+2\int_0 ^\frac{1}{\sqrt{r_0^2+l^2}}\,du\,\Big(1+\frac{l^2u^2}{2}\Big)\Big[\frac{1-(2\,m\,u-Q^2u^2)(1-\omega)}{1-2\,m\,u+(a^2+Q^2)u^2}\Big]\frac{1}{\sqrt{B(u)}}+\mathcal{O}(l^4)\nonumber 
\\&=&\hat{\alpha}_{KN}\Big|_{\sqrt{{r_0^2+l^2}}}+l^2\,\xi(m,a,\lambda,Q)+\mathcal{O}(l^4)\,,\label{eqq2}
\end{eqnarray}
where, $$\hat{\alpha}_{KN}\Big|_{\sqrt{{r_0^2+l^2}}}=-\pi+2\int_0 ^\frac{1}{\sqrt{r_0^2+l^2}}\,du\,\Big[\frac{1-(2\,m\,u-Q^2u^2)(1-\omega)}{1-2\,m\,u+(a^2+Q^2)u^2}\Big]\frac{1}{\sqrt{B(u)}}.$$
In, $l=0$ limit,  $\hat \alpha_{\textrm{KN}}$ reduces to the deflection angle for Kerr-Newman black hole as mentioned in (\ref{46s}) and 
\begin{eqnarray}
   {\xi}(m,a,\lambda,Q)&=& \int_0^{\frac{1}{\sqrt{r_0^2+l^2}}} u^2 \Bigg[\frac{1-(2mu-Q^2u^2)(1-\omega)}{1-2mu+(a^2+Q^2)u^2}\Bigg]\frac{1}{\sqrt{B(u)}}\,du\,,\label{412}
\end{eqnarray}
with $\omega=\frac{a}{\lambda}.$ We now rewrite $B(u)$ as,
\begin{eqnarray}
   B(u)= - Q^2(1-w)^2(u-u_1)(u-u_2)(u-u_3)(u-u_4)\,,\label{310}
\end{eqnarray}
where the roots are defined as,
\begin{eqnarray}
 u_1&=&\frac{X_1-2m-X_2}{4m\sqrt{r_0^2+l^2}}\,,\label{311}\\
 u_2&=&\frac{1}{\sqrt{r_0^2+l^2}}\,,\\
 u_3&=&\frac{X_1-2m+X_2}{4m\sqrt{r_0^2+l^2}}\,,\\
 u_4&=&\frac{2m}{Q^2}-\frac{X_1}{2m\sqrt{r_0^2+l^2}}\,.\label{314}
\end{eqnarray}
Again we apply the same strategy as sketched in the Appendix~(\ref{appendixA}) for the Kerr-Newman case. We choose the constants $X_1$ and $X_2$ in such a way, so that we can write down the roots in the following order $u_1<u_2<u_3<u_4$. Just like the Kerr-Newman case, $u_2,u_3,u_4$ turn out to be the positive roots while $u_1$ turns to be a negative root. To find out the roots, we need to substitute equations (\ref{311}) to (\ref{314}) into (\ref{310}) and compare it with the coefficients of $u^0,u^2,u^3,u^4$ in (\ref{43}) we will get \cite{Hsiao:2019ohy},
\begin{align}
\begin{split}
     & Q^2\Big[X_2^2-(X_1-2m)(X_1+6m)+4X_{1}^2\Big]=16m^2\sqrt{r_0^2+l^2}\Big(X_1-\sqrt{r_0^2+l^2}\frac{1+\omega}{1-\omega}\Big)\,,\\ &
     X_2^2-(X_1^2-2m)^2=\frac{8m(X_1-2m)(Q^2X_1-4m^2\sqrt{r_0^2+l^2})}{Q^2(X_1-2m)-4m^2\sqrt{r_0^2+l^2}}\,,\\ &
     \Big[X_2^2-(X_1^2-2m)^2\Big]\Bigg(\frac{1}{8m(r_0^2+l^2)^{\frac{3}{2}}}-\frac{Q^2X_1}{32m^3(r_0^2+l^2)^2}\Bigg)=\frac{1}{\lambda^2(1-\omega)^2}\,.\label{418}
     \end{split}
\end{align}
Exactly mimicking the same procedure as Sec~(\ref{3.3}) one can write down the expression of deflection angle upto $\mathcal{O}(l^2)$ and the correction term is given by,
\begin{align}
\begin{split}
  \textstyle{{\xi}(m,a,\lambda,Q)} & =
\textstyle{ \int_0^{u_2} du\,  u^2\Big[\frac{G_+}{u_+-u}+\frac{G_-}{u_--u}+\frac{G_{Q+}u}{u_+-u}+\frac{G_{Q-}u}{u_--u}\Big]\frac{1}{\sqrt{-Q^2(1-\omega)^2(u-u_1)(u-u_2)(u-u_3)(u-u_4)}}}\\ &
\textstyle{ = G_{+}\,g\, (u_+\,\Delta F+[(u_1-u_4)\Delta\Pi(\alpha^2)+u_4\Delta F]
     -u_+^2\frac{1}{(u_+-u_1)}[\frac{u_1-u_4}{u_+-u_4}\Delta\Pi(\alpha_{+3}^2)+\frac{u_+-u_1}{u_+-u_4}\Delta F])}\\ &
   \textstyle{+G_{-}\,g\,(u_-\,\Delta F+[(u_1-u_4)\Delta\Pi(\alpha^2))+u_4\Delta F]
     -u_{-}^2\frac{1}{(u_--u_1)}[\frac{u_1-u_4}{u_--u_4}\Delta\Pi(\alpha_{-3}^2)+\frac{u_--u_1}{u_--u_4}\Delta F])}\\ &
   \textstyle{+G_{Q+}\, g [(u_4^2-2u_+u_4+u_+^2-\frac{u_+^3}{u_+-u_4})\Delta F+(u_+(u_1-u_4)-\frac{u_+^3(u_1-u_4)}{(u_+-u_1)(u_+-u_4)})\Delta\Pi(\alpha_{+3}^2)}\\ & \textstyle{-2u_4(u_1-u_4)\chi_1(\alpha^2)-(u_1-u_4)^2\chi_2(\alpha^2)]+G_{Q-}\, g\,[(u_4^2-2u_-u_4+u_-^2-\frac{u_-^3}{u_--u_4}) \Delta F}\\ & \textstyle{+(u_-(u_1-u_4)-\frac{u_-^3(u_1-u_4)}{(u_--u_1)(u_--u_4)}) 
\Delta\Pi(\alpha_{-3}^2)-2u_4(u_1-u_4)\chi_1(\alpha^2)-(u_1-u_4)^2\chi_2(\alpha^2)]}\label{431}
 \end{split}
 \end{align}
 where,
\begin{align}
    \begin{split}
   & \Delta F=F(\psi_{0},k)-F(\frac{\pi}{2},k),
   \,\,\Delta\Pi(\alpha^2)=\Pi(\psi_{0},\alpha^2,k)-\Pi(\frac{\pi}{2},\alpha^2,k),\,\,\\&\Delta\Pi(\alpha_{\pm3}^2)=\Pi(\psi_{0},\alpha_{\pm3}^2,k)-\Pi(\frac{\pi}{2},\alpha_{\pm3}^2,k)\label{432}
    \end{split}
 \end{align}
 and, $\chi_1$ and $\chi_2$ is given by,
\begin{align}
\begin{split}
& \chi_1(\alpha^2)=\Pi(\frac{\pi}{2},\alpha^2,k)-\Pi(\psi_{0},\alpha^2,k)\,,\\ &
\chi_2(\alpha^2)=\frac{1}{2(\alpha^2-1)(k^2-\alpha^2)}\Big[\alpha^2\Big(E(\frac{\pi}{2},k)-E(\psi_{0},k)\Big)+(k^2-\alpha^2)\Big(F(\frac{\pi}{2},k)-F(\psi_{0},k)\Big)\,,\\ &
(2\alpha^2\,k^2+2\alpha^2-\alpha^4-3k^2)\Big(\Pi(\frac{\pi}{2},\alpha^2,k)\Big)-\Pi(\psi_{0},\alpha^2,k)\Big)\Big)-\frac{\alpha^4\,\sin\,\psi_{0}\sqrt{1-\sin^2\psi_{0}}\sqrt{1-k^2\,\sin^2\psi_{0}}}{1-\alpha^2 \,\sin^2\psi_{0}}\Big]
\end{split}
\end{align}
 and 
 \begin{align}
     \begin{split}
           & g=\frac{2}{\sqrt{Q^2(1-\omega)^2(u_4-u_2)(u_3-u_1)}},\,\, \alpha^2=\frac{u_1-u_2}{u_4-u_2}<0,\,\,
     \,\,
      \alpha_{\pm 3}^2=\alpha^2\frac{u_{\pm}-u_4}{u_{\pm}-u_1}\,,\\&
       k^2=\frac{(u_4-u_3)(u_2-u_1)}{(u_4-u_2)(u_3-u_1)},\,\,
       \psi_{0}=\arcsin\sqrt{\frac{(X_2+2m-X_1)[4m^2\sqrt{r_0^2+l^2}-Q^2(X_1+2m)]}{(X_2+6m-X_1)(4m^2\sqrt{r_0^2+l^2}-Q^2X_1)}}.
     \end{split}
 \end{align}
 $\Pi(\psi_{0},\alpha^2,k)$ and $\Pi(\frac{\pi}{2},\alpha^2,k)$ are the incomplete and complete elliptic integrals of the third kind, respectively. On the other hand, $F(\psi_{0},k)$ and $F(\frac{\pi}{2},k)$ are the  incomplete and complete first-kind elliptic integrals respectively. In addition, we have $E(\psi_{0},k)$ and $E(\frac{\pi}{2},k)$ in $\boldsymbol{V}_{1,2}$, which are incomplete and complete elliptic integrals of the second kind, respectively. The computation details are given in \cite{Ghosh:2022mka}. So just the like Kerr-Newman case, we  provide an analytical expression of equatorial deflection angle for Kerr-Newman black-bounce metric upto $\mathcal{O}(l^2)$ in this section. 
\subsection{Strong deflection analysis}
Following the same analysis given in Sec~(\ref{3.4}) we can find out the deflection angle of equatorial light rays in strong deflection limit. We will summarize the steps and results below:
\begin{itemize}
    \item First write down the deflection angle as follows,
    \begin{eqnarray}
          \phi(r_0)=\int_0^1\Bar{R}(z_2,r_0)\Bar{F}(z_2,r_0) dz_2\label{59}
    \end{eqnarray}
    with,
    \begin{eqnarray}
   \Bar{R}(z_2,r_0)&=&\frac{2(1-y_0)}{A'}\frac{\sqrt{\Bar{B}\Bar{A}_0}(\Bar{D}+2L\Bar{A})}{\sqrt{4\Bar{AC}^2+\Bar{CD}^2}}\\
   \Bar{F}(z_2,r_0)&=&\frac{1}{\sqrt{\frac{1}{\Bar{C}}(\Bar{CA}_0-\Bar{AC}_0+L(\Bar{AD}_0-\Bar{DA}_0)})}=\frac{1}{\sqrt{\Bar{H}}}\,,\label{H1}\\
   \Bar{H} & =&\frac{1}{\Bar{C}}(\Bar{CA}_0-\Bar{AC}_0+L(\Bar{AD}_0-\Bar{DA}_0))\,.\label{H}
\end{eqnarray}
The integral is potentially divergent at $r_0=r_c$.
\item Secondly, seperate the convergent and the divergent integral as,
\begin{eqnarray}
      \phi(r_0)=\phi_{\Bar{R}}(r_0)+\phi_{\Bar{F}}(r_0)
\end{eqnarray}
where,
\begin{eqnarray}\label{4.24}
    \phi_{\Bar{F}}(r_0)&=&\int_0^1\Bar{R}(z_2=0,r_c)\Bar{F}_0(z_2,r_0)dz_2\,.\label{513}\text{$\rightarrow$ the divergent part }\\
    \phi_{\Bar{R}}(r_0)&=& \int_0^1\Bar{R}(z_2,r_0)\Bar{F}(z_2,r_0) dz_2-\int_0^1\Bar{R}(z_2=0,r_c)\Bar{F}_0(z_2,r_0)dz_2\,.\text{$\rightarrow$ the finite part }\nonumber
\end{eqnarray}
\item Next find out the  critical turning point $r_c$ by solving the equation,
\begin{eqnarray}
     \Bar{A}_0\Bar{C}_0'-\Bar{A}_0'\Bar{C}_0-L\, (\Bar{A}_0\Bar{D}_0'-\Bar{A}_0'\Bar{D}_0)\Big|_{r_0=r_c}=0\,.
\end{eqnarray}
which gives,
\begin{align}
    \begin{split}
 &
 2 a^2 \left(Q^2-m \sqrt{l^2+r_c^2}\right)-2 a \left(Q^2-m \sqrt{l^2+r_c^2}\right) \sqrt{a^2-2 m \sqrt{l^2+r_c^2}+l^2+Q^2+r_c^2}+l^4\\ &
 +l^2 \left(-5 m \sqrt{l^2+r_c^2} +6 m^2+3 Q^2+2 r_c^2\right)
 -7 m Q^2 \sqrt{l^2+r_c^2}-5 m r_c^2 \sqrt{l^2+r_c^2}\\ &
 +6 m^2 r_c^2+2 Q^4+3 Q^2 r_c^2+r_c^4=0 
    \end{split}
\end{align}
We reproduce the results of \cite{Ghosh:2022mka} for the dependence of the critical turning point on different spacetime parameters ($a, Q,l$) in Fig~(\ref{Fig3}).
\begin{figure}[ht!]
		\minipage{0.33\textwidth}
\includegraphics[scale=0.60]{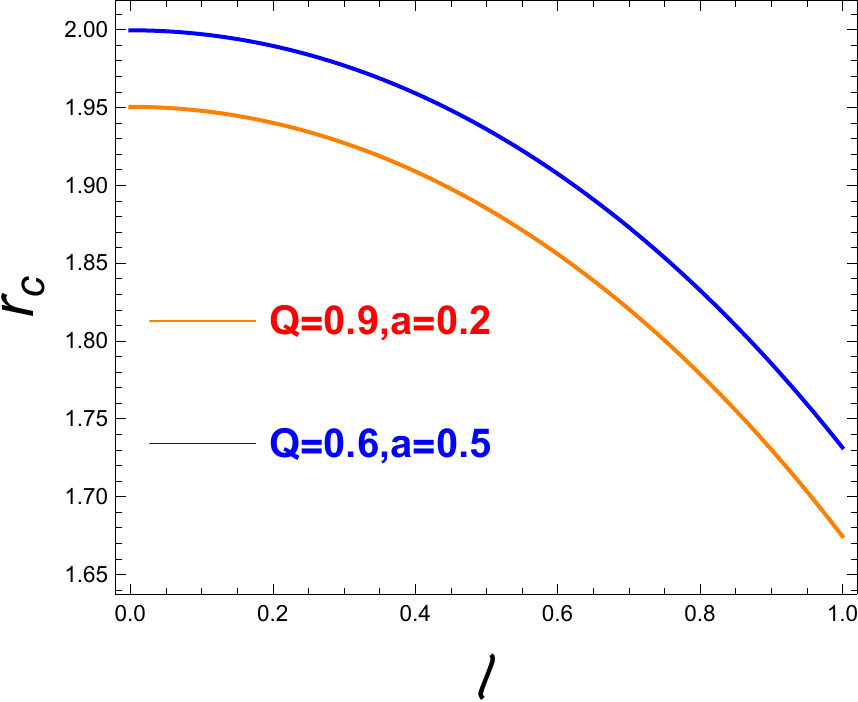}
	\endminipage\hfill
		\minipage{0.33\textwidth}
\includegraphics[scale=0.60]{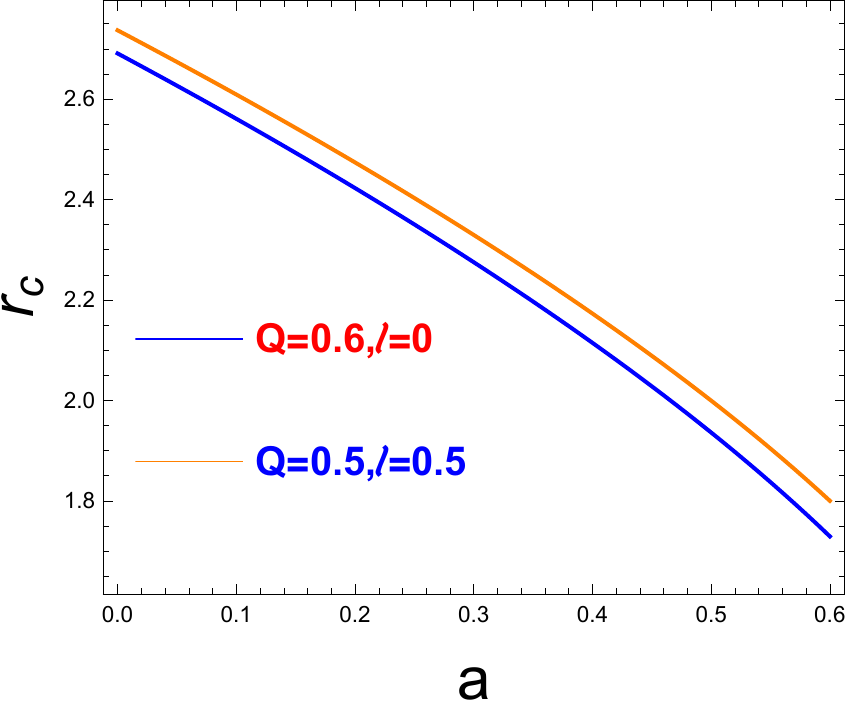}
\endminipage\hfill
			\minipage{0.33\textwidth}
\includegraphics[scale=0.60]{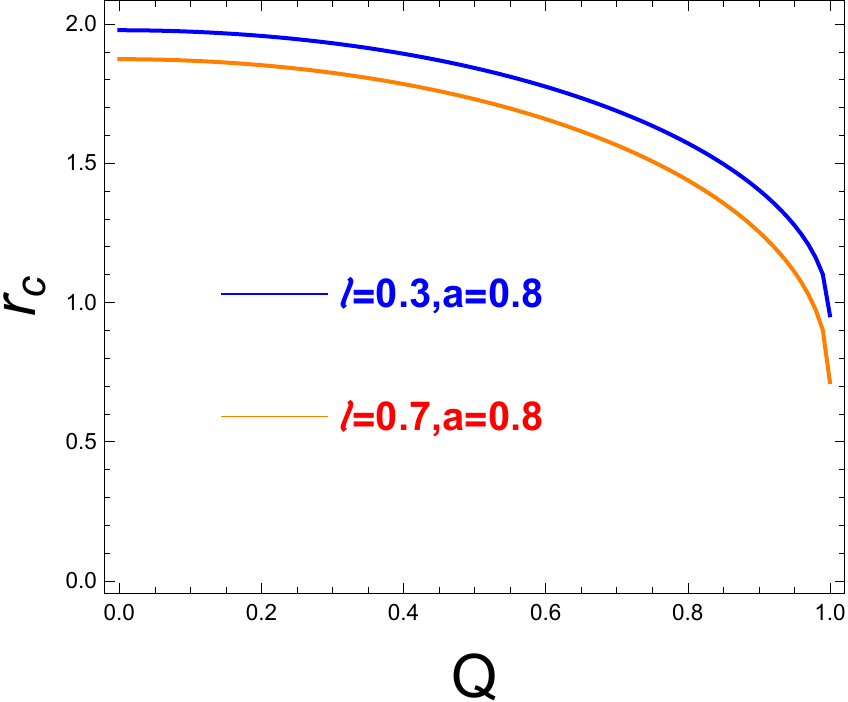}
\endminipage
\caption{Variation of radii of the photon sphere for charged, rotating Kerr-Newman black-bounce metrics for various values of $(a, Q, l)$. In the leftmost figure, we vary $l$ keeping $a$ and $Q$ fixed. In the middle figure we vary $a$ keeping $l$ and $Q$, and in the rightmost figure we vary $Q$ keeping $a$ and $l$ \cite{Ghosh:2022mka}.}
\label{Fig3}
\end{figure}
\item At the end calculate the integrals in (\ref{4.24}) at the limit $r_0\rightarrow r_c$ and we will find the logarthimic nature of the deflection angle as shown in Fig~(\ref{Fig4}). Again we reproduce the result of \cite{Ghosh:2022mka} here.
\begin{figure}[ht!]
\centering
\includegraphics[scale=0.63]{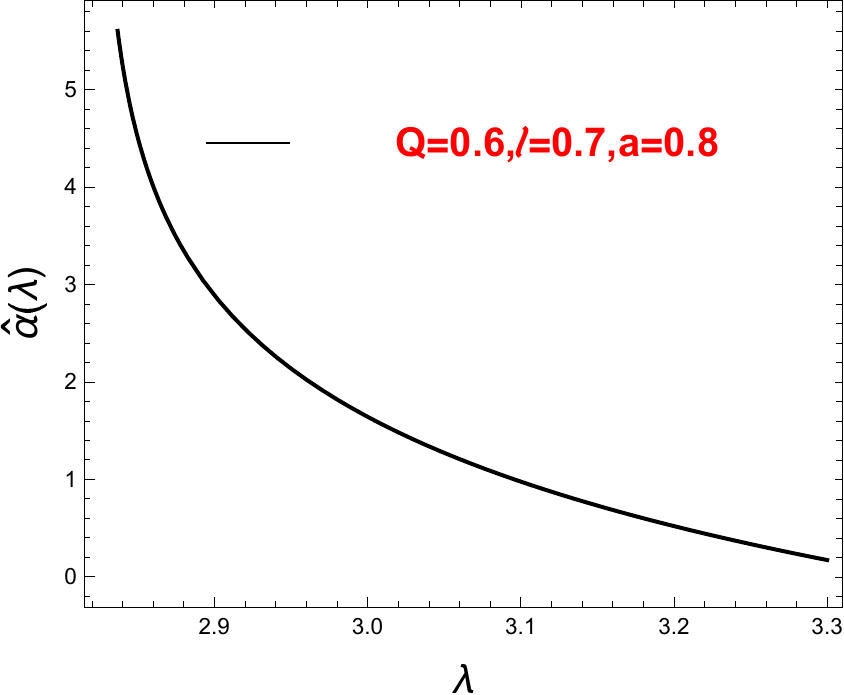}
\includegraphics[scale=0.63]{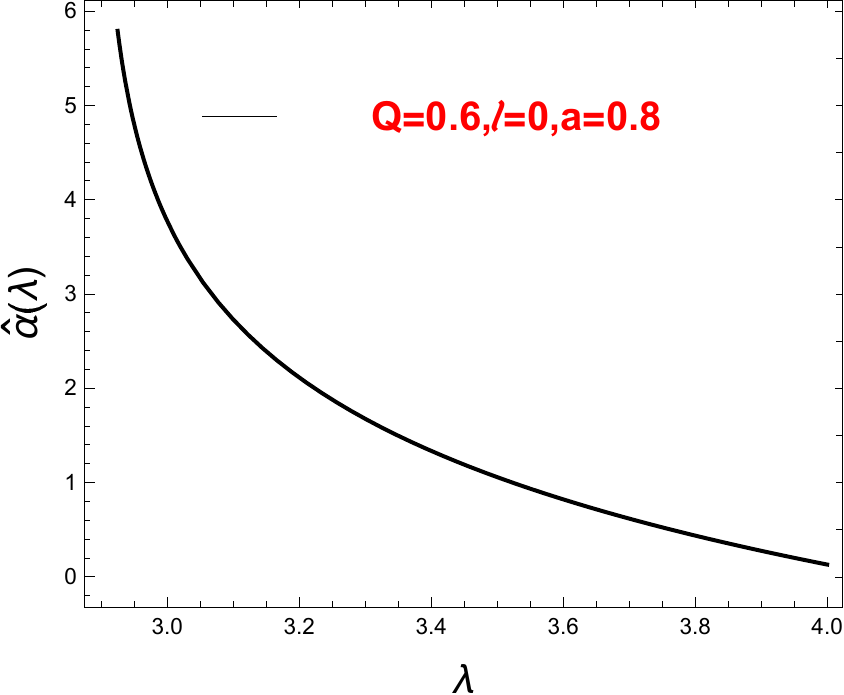}
\includegraphics[scale=0.63]{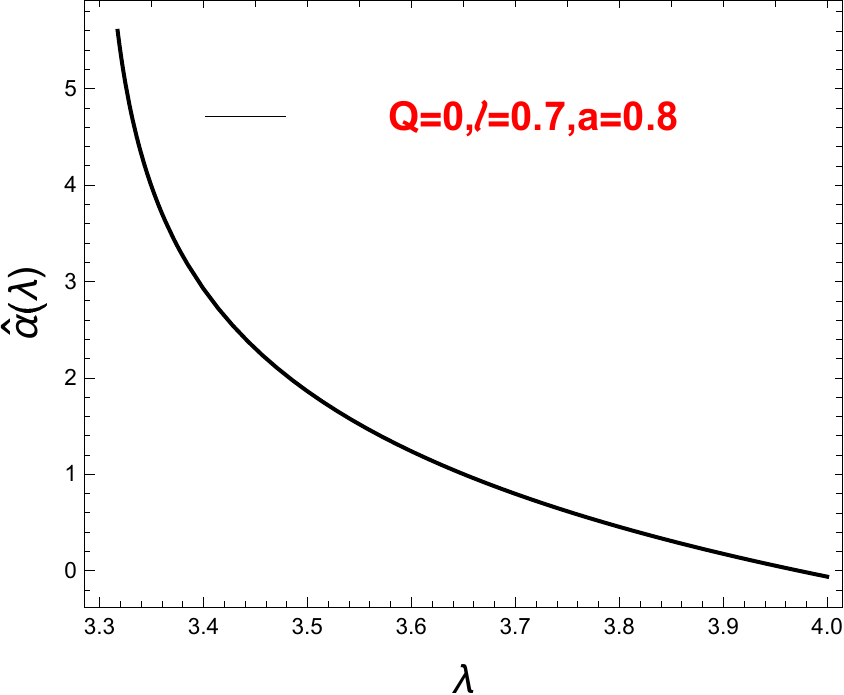}
\caption{Variation of the deflection angle $\hat{\alpha}$ with respect to the impact parameter $\lambda$ for fixed values of $Q,l,a$ \cite{Ghosh:2022mka}.}
\label{Fig4}
\end{figure}
\end{itemize}
\subsection{Observational signature in strong deflection limit}
Now we will discuss some observational consequences. First step for this is to relate the deflection angle and the angular radius of the Einstein ring. This is done by busing a lens equation. In this paper, we will use the following lens equation \cite{Bozza:2001xd},
\begin{eqnarray}
   \beta=\theta-\frac{\boldsymbol{D_{LS}}}{\boldsymbol{D_{OS}}}\Delta\alpha_n\label{61}
\end{eqnarray}
where $$\boldsymbol{D_{OS}}=\boldsymbol{D_{OL}}+\boldsymbol{D_{LS}}.$$ Also, $\beta$ is the angular separation between the source and the lens. $\boldsymbol{D}_{LS},\boldsymbol{D}_{OS},\boldsymbol{D}_{OL}$ are the distances between  lens to source, observer to source and observer to lens respectively. Finally, $$\Delta\alpha_n=\hat{\alpha}(\theta)-2n\pi.$$ 
The angular separation between the lens and the  $n^{th}$ image can be written as, $$\theta_n=\theta_n^0+\Delta \theta_n,$$ where $\theta_n^0$ is the angular separation between the $n$ image and {\bf the lens when the extra deflection angle ($\Delta\theta_n$) over $2n\pi$ is negligible.}
\begin{figure}[t!]
    \centering
 \includegraphics[scale=0.75]{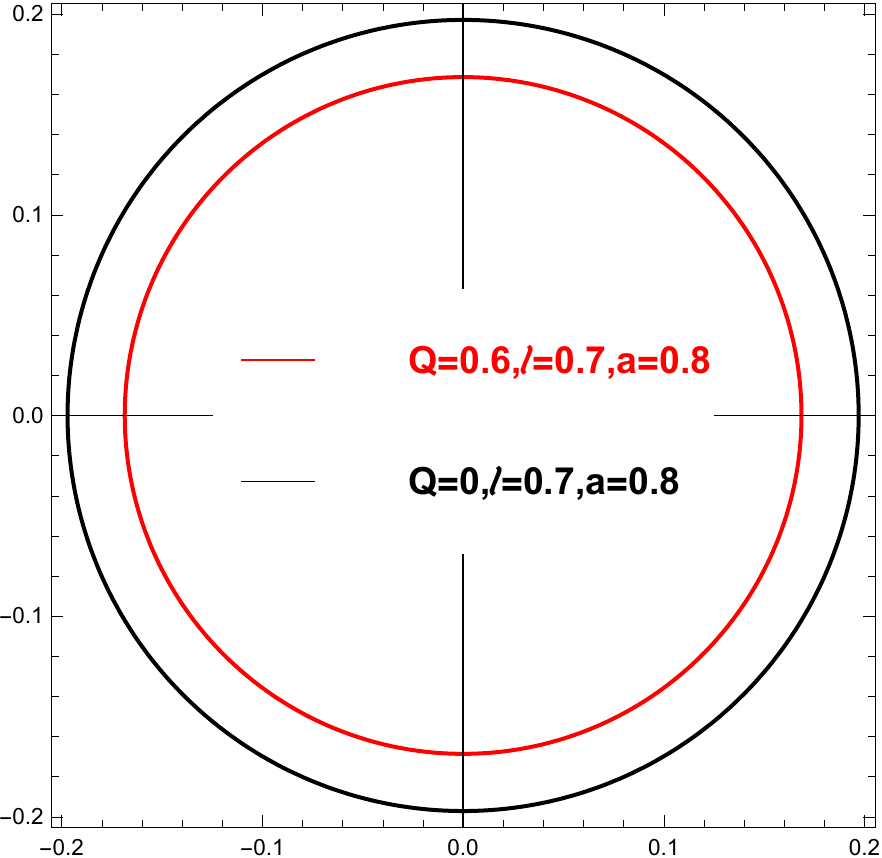}
 \includegraphics[scale=0.75]{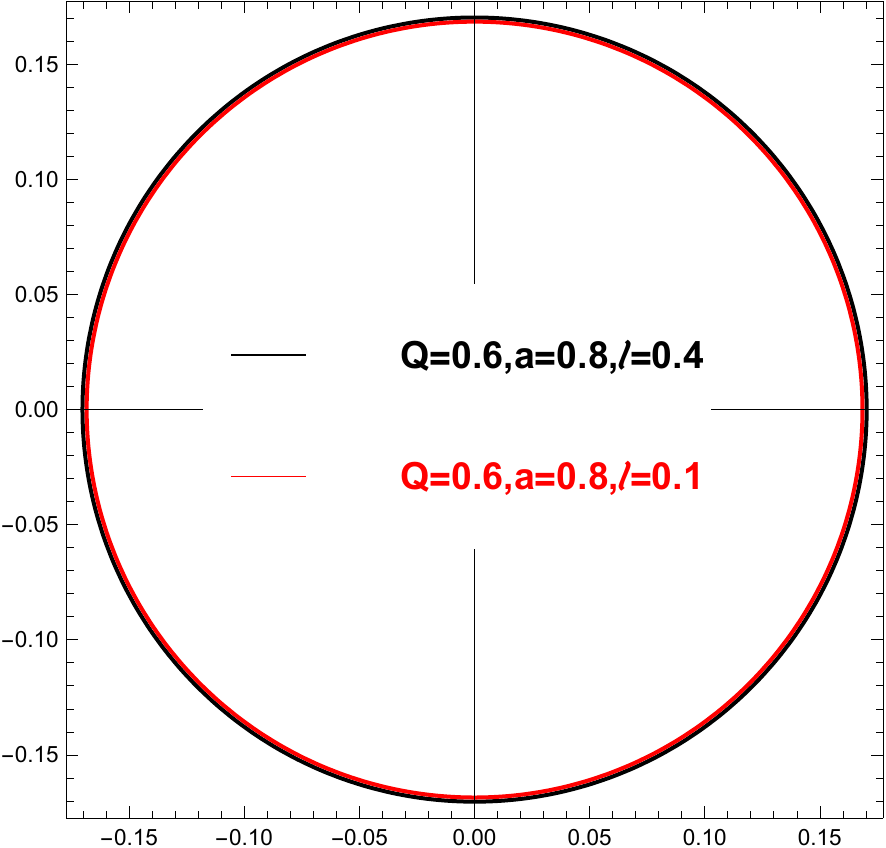}
\caption{Polar plots showing the angular radius $\theta_1$ for different $Q,l,a$ values \cite{Ghosh:2022mka}.}
    \label{Fig5}
\end{figure}

\par For the perfect alignment i.e, when $\beta=0$ and assuming $\Delta\theta_n<<\theta_n^0,$ the angular separation (angular radius) can be written as \cite{Bozza:2002af},
\begin{eqnarray}
   \theta_n^{Einstein}=\frac{\lambda_c}{\boldsymbol{D_{OL}}}\Big[1+\exp\Big(\frac{\Bar{b}}{\Bar{a}}-\frac{2n\pi}{\Bar{a}}\Big)\Big]\,. \label{neq1}
\end{eqnarray}
$n=1$ corresponds to the outermost Einstein ring. Following \cite{Ghosh:2022mka}, we plot some of these rings for different values of $Q,a$ and $l$ in Fig.~(\ref{Fig5}).\par
From the left-most plot in Fig.(\ref{Fig5})) we can conclude that as the charge of the black hole $Q$ increases (for fixed $l$ and $a$), the radii of the ring decreases. On the other hand, the effect of changing $l$ (for fixed $a$ and $Q$) on the ring radius is negligible. This is evident from the right-most plot in Fig. (\ref{Fig5}).\par
To make this observation more concrete, we carry out a detailed study, as shown in Tables (\ref{Tab1} and \ref{Tab2}). Some of the values provided in (\ref{Tab1} and \ref{Tab2}) are reproduced from \cite{Ghosh:2022mka}. We've listed some values regarding the representative percentage change  in the angular radius of the outermost Einstein rings with respect to $Q$ (for fixed $a$ and $l$) and $l$ (for fixed $a$ and $Q$) in Table~(\ref{Tab1}) and (\ref{Tab2}) respectively. These values corroborate perfectly the conclusion drawn above.\par

 \begin{table}[b!]
\centering
\begin{tabular}{|p{1.5cm}|p{0.75cm}|c|c|p{1.5cm}|p{0.75cm}|} 
 \hline
 \multicolumn{2}{|c}{Angular separation}&\multicolumn{1}{|c}{Percentage change}&\multicolumn{1}{|c}{Values of Charge}&\multicolumn{2}{|c|}{Fixed parameters} \\
 \hline
 $\theta_1^{(1)}(Q_2)$&$\theta_1^{(2)}(Q_1)$&$\Delta\theta =\frac{(\theta_1^{(2)}-\theta_1^{(1)})}{\theta_1^{(1)}}\times 100$ & ($Q_1,Q_2$) & $l$ & $a$ \\ 
 \hline
 0.1686 & 0.1970 &16.9013 & (0, 0.6) & 0.7 & 0.8 \\ 
 \hline
0.18548 & 0.21590 & 16.399 & (0.6,0.8) & 0.4 & 0.5 \\
 \hline
 0.22691 & 0.24236 & 6.8088 & (0.45,0.65) & 0.4 & 0.5\\
 \hline
\end{tabular}
 \caption{Percentage change of the angular radius of first Einstein ring for different values of charge $Q$ for fixed $a$ and $l.$ Some of the  numerical values presented here are reproduced from \cite{Ghosh:2022mka}.}
\label{Tab1}
\end{table}
\begin{table}[t!]
\centering
\begin{tabular}{|p{1.5cm}|p{0.75cm}|c|c|} 
 \hline
 \multicolumn{2}{|c}{Angular separation}&\multicolumn{1}{|c}{Percentage change}&\multicolumn{1}{|c|}{\makecell{Values of\\ regularization\\ parameter}} \\
 \hline
 $\theta_1^{(1)}(l_1)$&$\theta_1^{(2)}(l_2)$&$\Delta\theta =\frac{(\theta_1^{(2)}-\theta_1^{(1)})}{\theta_1^{(1)}}\times 100$ & $(l_1,l_2)$ \\ 
 \hline
 0.16856 & 0.1703 & 1.07059 & (0.2,0.4)  \\ 
 \hline
 0.16756 & 0.16760 & 0.023872 & (0.25,0.75) \\
 \hline
 0.166678 & 0.166680 & 0.001199 & (0.15,0.25)\\
 \hline
\end{tabular}
 \caption{Percentage change of the angular radius of first Einstein ring for different values of regularisation parameter $l$ for  $a=0.8$ and $Q=0.6.$  Some of the numerical values presented here are reproduced from \cite{Ghosh:2022mka}.}
\label{Tab2}
\end{table}
Before closing this section, a few comments regarding possible avenues to constraints on the spacetime parameters should be in order utilizing observational data. One of the ways is to look into the ratio of mass to distance, the mass being the mass of the central object (e.g. Sagittarius $A^*$), which is around $4.4\times 10^{6}M_{\odot}$ and its distance being $8.5$ kpc, the ratio turns out to be around $2.4734\times 10^{-11}$. One can use this data to give the angular position of the relativistic images and the angular separation between the two Einstein rings. On the other hand, we can compute the angular separation of two successive Einstein ring from (\ref{neq1}) for different values of $n$. Then one can utilize it to build a parameter space for the spacetime parameters. Interested readers are referred to, e.g. \cite{Wang:2016paq} for a more comprehensive discussion. In future detections, if one can better resolve the angular separation of various Einstein rings, we will get better constraints on the charge of the underlying black-bounce metric.
\par
\subsection{Results for non equatorial lensing: caustic points}\label{Sec4.5}
In Sec.~(\ref{sec3.5}) we discussed about the non equatorial lensing for Kerr-Newman black hole. For this case also the analysis would be same.  Only difference is that the scaling factor $\omega(r)$  will be function  of the regularisation parameter $l$ apart from $Q$ and $a.$
\begin{eqnarray}
 \omega(r)= \bar{\lambda}\,\frac{a^2+\sqrt{r^2+l^2}(\sqrt{r^2+l^2}-2)}{\sqrt{r^2+l^2}\Big(2\,a+\lambda(\sqrt{r^2+l^2}-2)\Big)+Q^2(\lambda-a)}
\end{eqnarray}
with, $\bar{\lambda}=\sqrt{\lambda^2-a^2}$.

\begin{figure}[htb!]
    \centering
 \includegraphics[scale=0.50]{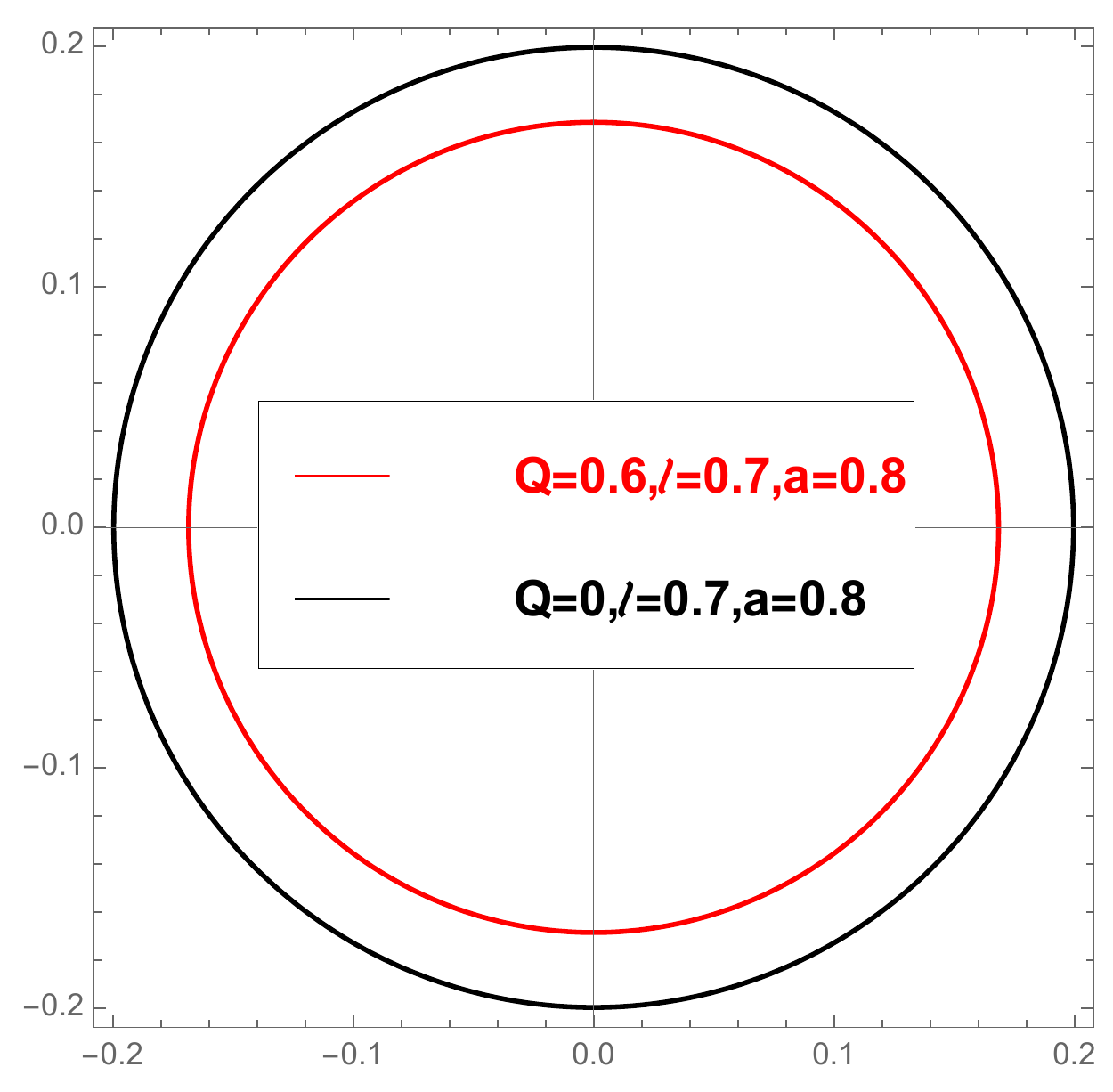}
 \includegraphics[scale=0.50]{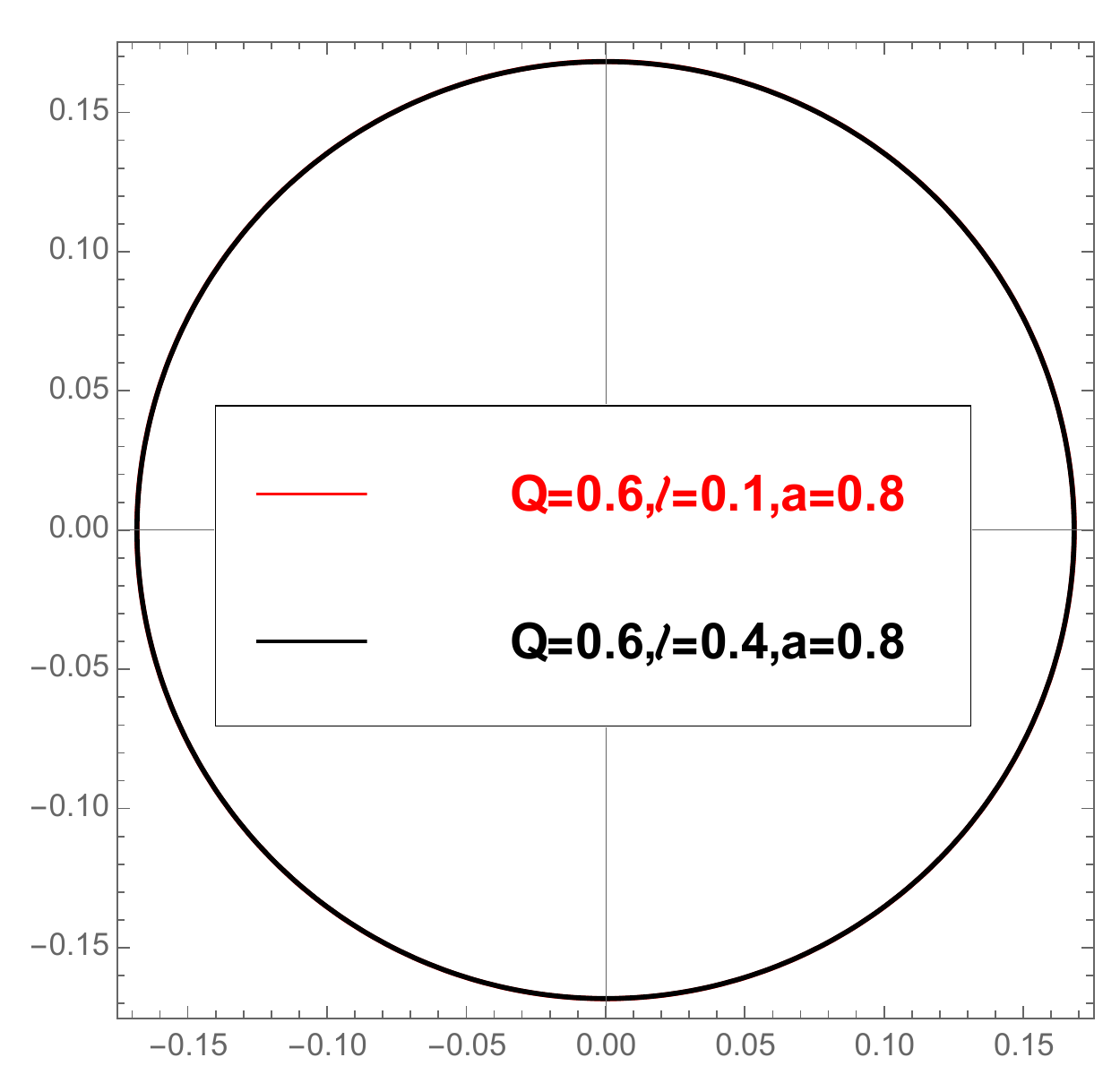}
\caption{Polar plots showing the angular radius $\theta_1$ for different $Q,l,a$ values \cite{Ghosh:2022mka}.}
    \label{Fig6}
    \end{figure}
Next we show how the angular radius of first Einstein ring  depends on $Q$ and $l$ in Fig~(\ref{Fig6}) following \cite{Ghosh:2022mka}. It is evident from the Fig~(\ref{Fig6}), that the  dependence of the angular radius on charge parameter ($Q$) of the black bounce metric is significaantly more than the dependence on the regularisation parameter ($l$) similar to the  case of equatorial lensing.\par
 \begin{table}[]
\centering
\begin{tabular}{|p{2cm}|p{0.75cm}|c|c|p{1.5cm}|} 
 
 \hline
 ($l,a$)&$Q$&$ \bar{\sigma_2}$ \\ 
 \hline
 (0.4,0.5) & 0.55 & -6.28\\
 \hline
  (0.4,0.5) & 0.65 & -4.98\\
  \hline 
  (0.4,0.5) & 0.75 & -0.137\\
  \hline
   (0.4,-0.5) & 0.5 & -5.69\\
   \hline
   (0.4,-0.5) & 0.6 & -5.84\\
   \hline
   (0.4,-0.5) & 0.7 & -5.90\\
   \hline
\end{tabular}
 \caption{Angular position of the second caustic point for different values of $Q$. The first three entities are for direct photon and the last three are for retrograde photon. Some of the numerical values presented here is reproduced from \cite{Ghosh:2022mka}}
\label{Tab5}
\end{table}
In Table ~(\ref{Tab5}), we investigate the variation of the second caustic point with respect to $Q$ for fixed $l$ and $a$.
\begin{table}[htb!]
\centering
\begin{tabular}{|p{2cm}|p{0.75cm}|c|c|p{1.5cm}|} 
 
 \hline
 ($Q,a$)&$l$&$ \bar{\sigma_2}$ \\ 
 \hline
 (0.5,0.5) & 0.45 & -6.646\\
 \hline
  (0.5,0.5) & 0.55 & -6.702\\
  \hline 
  (0.5,0.5) & 0.65 & -6.79\\
  \hline
   (0.5,-0.5) & 0.7 & -5.92\\
   \hline
   (0.5,-0.5) & 0.8 & -5.98\\
   \hline
   (0.5,-0.5) & 1.0 & -6.20\\
   \hline
\end{tabular}
 \caption{Angular position of the second caustic point for different values of $l$. The first three entities are for direct photons, and the last three are for retrograde photons. Some of the numerical values presented here is reproduced from \cite{Ghosh:2022mka}.}
\label{Tab6}
\end{table}
Before closing this section, we  further investigate the variation of the second caustic point with respect to $l$ for fixed $Q$ and $a$. It is demonstrated in the Table ~(\ref{Tab6}). 
Comparing the values in Table ~(\ref{Tab5}) and (\ref{Tab6}) we can conclude that the change of caustic points for different $Q$ and different $l$ is not that robust \cite{Ghosh:2022mka}.

\section{Conclusion}\label{sec9}
As mentioned earlier, gravitational lensing studies provide an excellent tool to provide insight into the structure of spacetime itself. In light of these advantages, this review provides a brief tour of the analytic methods used to calculate observables which can be measured. First we review some facts about lensing in Schwarzschild geometry, and observed the following points listed below.
\begin{itemize}
    \item The geodesics in this geometry are studied. As expected, owing to the symmetry of the spacetime itself, we have two constants of motion. Looking closely into the geodesic structures, we can find the location of the photon rings around them. Not only that, one can see that these photon rings are located at $3M$ are unstable, and one can seek out some non-trivial physics once you deviate infinitesimally from this range.
    \item After we have understood how geodesics behave in such a geometry, we can calculate quantities using the equations at hand. A comprehensive and self-explanatory calculation is provided which gives an estimate of the deflection angle in terms of the central massive object responsible for this deflection. The contribution of this central massive object in the formula is through the mass of the object. There's also a contribution from a $r_{min}$ term in the denominator which indirectly also depends on the metric structure around the central massive object. We have listed also listed down some salient features related to the calculation in the bullet points below equation (\ref{rev7}).
\end{itemize}
After giving a brief overview of the time delay suffered by these geodesics and also going on to calculate the diameter of the Einstein ring in this setup, we move on to a more general spacetime having extra rotation parameters. As expected, all the above observables will have a non-trivial rotation parameter-dependent term. The calculations are all in the strong deflection limit and analytical. We also consider a more general class of proposed solutions called black-bounce spacetimes, where there is an extra parameter involved as a deviation of the already known solutions in GR. We list the salient features of our findings below:
\begin{itemize}
    \item We presented a method of calculating the deflection angle analytically by doing a perturbation in $l.$ The results are given in terms of elliptic integrals of various kinds. Also, we have restricted ourselves to the equatorial plane while doing this analysis. We observe that for non-zero $l,$, the value of the deflection angle for a fixed impact parameter decreases. Our calculation provides a general methodology to compute deflection angles analytically in a perturbative series.  In future it will be interesting to go beyond this small $l$ expansion. This will require a thorough numerical analysis. 
    
    \item Next, we study the strong deflection limit of the equatorial deflection angle. This has direct observational implications because it provides information about the Einstein rings. We can conclude that the effect of the charge ($Q$) on the size of the Einstein ring is much more pronounced than the effect of the regularisation parameter ($l$) for a fixed value of spin parameter $a$. We discovered that decreasing the charge ($Q$) considerably increases the ring's size. This observation remained the same even when we computed the ring's radius for a small polar inclination. 
\item Furthermore, we extend our analysis for non-equatorial lensing. This enables us to compute the location of the caustic points. We again observed that the effect of the charge ($Q$) on the position of caustic points is more pronounced than the regularisation parameter ($l$). Again one can apply this method to different black hole spacetime to get an exact analytical expression. The study of non-equatorial lensing presented in this review assumes a small inclination angle. Again, it will be interesting to go beyond this regime. 
\end{itemize}
\par

Another interesting direction we could not include in this review  is the analysis of the structure of the shadow. Interested readers are referred to this review \cite{Perlick:2021aok} for more details about this topic. Analysis of the shadow structure complements the analysis of the Einstein ring, which is presented here. It provides further constraints on the different black hole parameters and various theories of gravity. Interested readers are again referred to \cite{Perlick:2021aok} for relevant references for this. More analytical works on gravitational lensing has been done in \cite{DeFalco:2016yox}.\par
The analysis of the deflection angle presented in this review can straightforwardly be repeated to other black-bounce spacetimes, for example, \cite{Barrientos:2022avi}. Finally, there are several other avenues which have been pursued recently in the context of strong deflection of light rays. One such thing is the study of multilevel images. This helps one to predict how much resolution is required to distinguish between different Einstein rings. One important aspect is also to study the two-point correlation function of intensity fluctuations on the photon ring, which result from the photon travelling through several orbits around the central object. This plays a significant role from the perspective of image analysis. This black-bounce metric might be subject to these studies along the lines of \cite{Hadar:2020fda}. These investigations will support our efforts to communicate with plausible astrophysical settings.
\section*{Acknowledgements}

Research of A.C. is supported by the Prime Minister's Research Fellowship (PMRF-192002-1174) of Government of India. A.B is supported by Mathematical Research Impact Centric Support Grant (MTR/2021/000490) by the Department of Science and Technology Science and Engineering Research Board (India) and Relevant Research Project grant\\ (202011BRE03RP06633-BRNS) by the Board Of Research In Nuclear Sciences (BRNS), Department of Atomic Energy (DAE), India. 

\appendix
\section{Details of the computation of the deflection angle for Kerr-Newman Spacetime} \label{appendixA}
 Here we give details of computing the integral mentioned in (\ref{46s}).
First we rewrite $B(u)$ in the factorized form.
\begin{eqnarray}
   B(u)= - Q^2(1-w)^2(u-u_1)(u-u_2)(u-u_3)(u-u_4)\,,\label{310ss}
\end{eqnarray}
where,
\begin{eqnarray}
 u_1&=&\frac{X_1-2m-X_2}{4m{r_0}}\,,\label{311s}\\
u_2&=&\frac{1}{{r_0}}\,,\\
 u_3&=&\frac{X_1-2m+X_2}{4m r_0}\,,\\
 u_4&=&\frac{2m}{Q^2}-\frac{X_1}{2m r_0}\,.\label{314s}
\end{eqnarray}
Then by choosing the constants $X_1$ and $X_2$, we can write down the roots in the following order: $u_1<u_2<u_3<u_4$. Here, the positive roots are $u_2,u_3,u_4$, while the negative root is $u_1$. Then we substitute equations (\ref{311s}) to (\ref{314s}) into (\ref{310ss}) and compare it with the coefficients of $u_0,u_2,u_3,u_4$ in (\ref{43}). This way we can eventually extract the roots. That gives \cite{Hsiao:2019ohy},
\begin{align}
\begin{split}
     & Q^2\Big[X_2^2-(X_1-2m)(X_1+6m)+4X_{1}^2\Big]=16m^2 r_0\Big(X_1-r_0\frac{1+\omega}{1-\omega}\Big)\,,\\ &
     X_2^2-(X_1^2-2m)^2=\frac{8m(X_1-2m)(Q^2X_1-4m^2r_0)}{Q^2(X_1-2m)-4m^2r_0}\,,\\ &
     \Big[X_2^2-(X_1^2-2m)^2\Big]\Bigg(\frac{1}{8m r_0^3}-\frac{Q^2X_1}{32m^3 r_0^4}\Bigg)=\frac{1}{\lambda^2(1-\omega)^2}\,.\label{418s}
     \end{split}
\end{align}
Combining the first and second equation of (\ref{418s}) we get the equation for $X_1$, which is given by,
\begin{align}
\begin{split}
   &  \frac{Q^2}{2m}X_1^3-\Big(Q^2+4mr_0\Big)X_1^2+\Bigg(4m^2r_0+2mQ^2+\frac{8m^3r_0^2}{Q^2}+\frac{2mr_0^2(1+\omega)}{(1-\omega)}\Bigg)X_1\\ &
   =4\,m^2\,Q^2+\frac{4m^2 r_0^2(1+\omega)}{(1-\omega)}+\frac{8m^2 r_0^3(1+\omega)}{Q^2(1-\omega)}\,.\label{419s}
\end{split}
\end{align}
The equation (\ref{419s}) can be exactly solved. The positive real root $X_1(m,\omega,Q,r_0)$ turns out to be the one mentioned in (\ref{equ21s}) of the Sec.~(\ref{3.3}). 
From (\ref{equ21}), one can easily check that, when $Q$ becomes zero,  $\delta$ becomes $\pi.$ In that limit, (\ref{equ21s}) can be written as,
\begin{eqnarray}
    X_1(m,\omega,Q=0,r_0)=r_0\frac{1+\omega}{1-\omega}\,.
\end{eqnarray}
It then exactly reproduces the result for the Kerr black hole \cite{Iyer:2009wa}.
Furthermore, we can reproduce the result for the Reissner-Nordstrom black hole in the limit,  $a=0$, and the root of the equation (\ref{419s}) then reduce to,
\begin{equation}
    X_1(m,Q,\omega=1,r_c)=2m\Big(\frac{2mr_0}{Q^2}-1\Big)\Big|_{r_c}\label{320}
\end{equation}
where $r_c$ is defined in (\ref{33})
and the result (\ref{320}) matches with the result of \cite{Hsiao:2019ohy}.\par
Now we try to factorize the remaining part of the integrand of (\ref{46s}). We get, 

\begin{eqnarray}
\displaystyle{\frac{1-2mu(1-\omega)+Q^2u^2(1-\omega)}{1-2mu+(a^2+Q^2)u^2} =\frac{G_+}{u_+-u}+\frac{G_-}{u_--u}+\frac{G_{Q+}\,u}{u_+-u}+\frac{G_{Q-}\,u}{u_- -u}}\label{equ25}
\end{eqnarray}
where,
\begin{eqnarray} \label{upos}
 u_{\pm}=\frac{m\pm\sqrt{m^2-(a^2+Q^2)}}{a^2+Q^2}\,.
\end{eqnarray}
and 
$G_-,G_+,G_{Q+},G_{Q-}$  are defined in (\ref{neweqs}) of the Sec.~(\ref{3.3}). Also, we can easily see from (\ref{upos}) $u_\pm$ are positive.


\bibliographystyle{utphysmodb}
\bibliography{refs}

\end{document}